\documentclass{aa}

\usepackage[utf8]{inputenc}
\usepackage[T1]{fontenc}
\usepackage{lmodern}
\usepackage{microtype}
\usepackage{float}
\usepackage{booktabs}
\usepackage{blindtext}
\usepackage{siunitx}
\usepackage{amssymb}
\usepackage[flushleft]{threeparttable}
\usepackage{textcomp,gensymb}
\usepackage{graphicx}
\usepackage{txfonts}
\usepackage{hyperref}
\hypersetup{colorlinks,linkcolor={blue},citecolor={blue},urlcolor={blue}}

\usepackage{lscape}
\usepackage{graphicx}
\usepackage{array,multirow}
\usepackage{multirow}
\usepackage{txfonts}
\usepackage{gensymb}
\usepackage{natbib}
\bibpunct{(}{)}{;}{a}{}{,} % to follow the A&A style
\usepackage{color}
\usepackage{xcolor}
\definecolor{mhi}{rgb}{0.6,0.0,0.6}

\colorlet{avi}{green!0!orange!100!}

\newcommand{\kms}{\,${\rm km\,s^{-1}}$}

\begin{document} 

\title{The Fornax Cluster VLT Spectroscopic Survey III -
Kinematical characterisation of globular clusters across the Fornax galaxy cluster}

\titlerunning{Globular cluster kinematics in the Fornax cluster}

\author{Avinash Chaturvedi  \inst{\ref{eso}, \ref{lmu}}
\and Michael Hilker     \inst{\ref{eso}}
\and Michele Cantiello \inst{\ref{inafteramo}}
\and Nicola R. Napolitano \inst{\ref{nichina}, \ref{inafnapoli}}
\and Glenn van de Ven \inst{\ref{vienna}}
\and Chiara Spiniello \inst {\ref{oxf}, \ref{inafnapoli}}
\and Katja Fahrion \inst{\ref{eso}}
\and Maurizio Paolillo \inst{\ref{uninap}, \ref{inafnapoli}}
\and Massimiliano Gatto \inst{\ref{uninap}, \ref{inafnapoli}}
\and Thomas Puzia \inst{\ref{puzia}}}

\institute{European Southern Observatory, Karl-Schwarzschild-Stra\ss{}e 2,
85748 Garching, Germany  \label{eso}
\email{avinash.chaturvedi@eso.org} 
\and
Ludwig Maximilian Universitat, Professor-Huber-Platz 2, D-80539 Munich, Germany \label{lmu}
\and
INAF - Astronomical Observatory of Abruzzo, Via Maggini, 
64100 -Teramo, Italy  \label{inafteramo}
\and
School of Physics and Astronomy, Sun Yat-sen University, Zhuhai Campus, 2 Daxue Road, Xiangzhou District, Zhuhai, P. R. China \label{nichina}
\and
INAF - Osservatorio Astronomico di Capodimonte, Salita Moiariello, 
16, 80131 - Napoli, Italy  \label{inafnapoli}
\and
Department of Astrophysics, University of Vienna,
Turkenschanzstra\ss{}e 17, 1180 Wien, Austria \label{vienna}
\and
Sub-Dep. of Astrophysics, Dep. of Physics, University of Oxford, Denys Wilkinson Building, Keble Road, Oxford OX1 3RH, UK
\label{oxf}
\and
University of Naples Federico II, C.U. Monte Sant’Angelo, Via Cinthia, 80126 Naples, Italy\label{uninap}
\and
Institute of Astrophysics, Pontificia Universidad Catolica de Chile, Avenida Vicuna Mackenna 4860, 
Macul, 7820436, Santiago, Chile \label{puzia}}

\date{Received 17 May 2021 / Accepted 16 September 2021}

% \abstract{}{}{}{}{} 
% 5 {} token are mandatory
 
 \abstract
  {The Fornax cluster provides an unparalleled opportunity to investigate in detail the formation and evolution of early-type galaxies in a dense environment. We aim at kinematically characterizing photometrically detected globular cluster (GC) candidates in the core of the cluster. We used the VLT/VIMOS spectroscopic data from the FVSS survey in the Fornax cluster, covering one square degree around the central massive galaxy NGC\,1399. We confirmed a total of 777 GCs, almost doubling the previously detected GCs, using the same dataset by \cite{Pota2018}. Combined with previous literature radial velocity measurements of GCs in Fornax, we compiled the most extensive spectroscopic GC sample of 2341 objects in this environment. We found that red GCs are mostly concentrated around major galaxies, while blue GCs are kinematically irregular and are widely spread throughout the core region of the cluster. The velocity dispersion profiles of blue and red GCs show a quite distinct behaviour. Blue GCs exhibit a sharp increase in the velocity dispersion profile from 250 to 400\kms\ within 5 arcminutes ($\sim$29 kpc / $\sim$1\,$r_{\rm eff}$ of NGC\,1399) from the central galaxy. The velocity dispersion profile of red GCs follows a constant value in between 200-300\kms\ until 8 arcminutes ($\sim$46 kpc/ $\sim$1.6\,$r_{\rm eff}$), and then rises to 350\kms\ at 10 arcminutes ($\sim$58 kpc/ $\sim$2\,$r_{\rm eff}$). Beyond 10 arcminutes and out to 40 arcminutes ($\sim$230 kpc/ $\sim$8\,$r_{\rm eff}$), blue and red GCs show a constant velocity dispersion of 300$\pm$50\kms, indicating that both GC populations are tracing the cluster potential. We have kinematically confirmed and characterized the previously photometrically discovered overdensities of intra-cluster GCs. We found that those substructured intra-cluster regions in Fornax are dominated mostly by blue GCs.}


\maketitle
%-------------------------------------------------------------------

\section{Introduction}

Understanding the assembly of galaxy clusters provides valuable insight into various aspects of cosmology, like galaxy evolution and formation, gravitational structure formation, intergalactic medium physics, etc. Galaxy clusters are the largest gravitationally bound systems, whose assembly is driven by early mergers of massive galaxies embedded in big dark matter (DM) halos and sequential accretion of galaxy groups \citep[e.g.][]{Kravtsov2012}. During their growth, various physical processes act on the cluster galaxies, like tidal disturbances, ram pressure stripping, secular evolution, and gas accretion, which all contribute in shaping their luminous and dark matter distributions \citep{Kravtsov2012, Duc2011, Amorisco2019}.
Semi-analytic models of galaxy formation and evolution, combined with cosmological N-body simulations of DM halos in the $\Lambda$CDM framework, have shown that the amount of substructures in stellar halos and their dynamics directly probe two fundamental aspects of galaxy formation: the hierarchical assembly of massive galaxies and their DM halos \citep{Cooper2013, Pillepich2015}. 

The interaction processes leave dynamical imprints on the stellar populations of galaxies. In particular, tidal features are preserved and easily identified in the outer halos of galaxies, where the dynamical timescales are longer than in the inner parts 
\citep[e.g.][]{Napolitano2003}. Observed disturbances include stellar streams and tidal structures in phase space \citep{Romanowsky2012, Coccato2013, Longobardi2015}. Therefore, stellar halos are crucial in understanding the formation and evolution of galaxies.
 
Due to the low surface brightness of the outer halos of galaxies, kinematical details from integrated light at large radii are mostly inaccessible with current spectrographs. However, discrete tracers like globular clusters and planetary nebulae (PNe) play a significant role in learning about the halos kinematics. These are bright sources that are easily detectable in the outskirts of galaxies \citep{Dolfi2021, Longobardi2015, Longobardi2018, Hartke2018}.PNe represent a post-main sequence evolutionary stage of stars and are mainly associated with the stellar populations and integrated light of the galaxies \citep{Douglas2007, Coccato2009, Napolitano2011, Spiniello2018}. Globular clusters (GCs), on the other hand, are massive, compact, and mostly old star clusters, found in almost all major types of galaxies \citep[e.g.][]{Brodie2006}. Observations have shown that GCs exist in two major sub-populations: red (metal-rich) GCs and blue (metal-poor) GCs. The red GCs are found to have radial number density profiles similar to the integrated light of their host galaxies, while blue GCs are spatially more extended into the intergalactic and intra-cluster regions and trace the metal-poor component of the stellar halos \citep{Schuberth2010, Hilker2015, Cantiello2018, Pota2018}.

These two GC sub-populations also show different kinematical characteristics. In most cases, the red GCs follow the stellar population kinematics of their parent galaxies, whereas blue GCs show a more erratic and complex kinematic behaviour \citep{Schuberth2010, Coccato2013, Napolitano2014, Cantiello2018, Pota2018}. The GCs' colour bi-modality is mainly associated with their bimodal metallicity distribution, although the relation between colour and metallicity is not entirely linear \citep{Cantiello2014, Katja_nlm_2020}. The colour bi-modality and distinct kinematical behaviour have been explained as the result of a two-stage formation scenario for massive galaxies \citep{Ashman1992, Kundu2001, Peng2006}. Cosmological simulations suggest that massive early-type galaxies grow and evolve in these two stages: first, rapidly with a high star formation rate and early compact mergers (in-situ), and later through the continuous accretion of smaller systems that build up the extended halo populations.

The red, metal-rich GCs are thought to have formed during the in-situ star formation process, whereas the blue, metal-poor GCs are added to the system via the accretion of low mass objects, like dwarf galaxies \citep{Forbes1997, Cote1997, Hilker1999b, Kravtsov2006, Tonini2013, Forbes2018}. Various studies of GC populations in galaxy clusters revealed that there exist so-called intra-cluster GCs (ICGCs), which are not bound to any individual galaxy \citep{Williams2007, Bergond2007, Bassino2006, Schuberth2008, Peng2011, AlamoMartnez2013}. The ICGCs might represent the first GCs formed in a cluster potential or could be tidally released GCs from multiple galaxy interactions \citep{White1987, West1995, Yahagi2005, Madrid2018, Harris2020}.
Although their formation mechanism is still debated \citep{Ramos2018}, their kinematics add additional constraints on the accretion and assembly history of their parent galaxy cluster. 

The above-mentioned properties of GCs make them privileged discrete tracers for the dynamical study of individual galaxies, as well as the mass assembly of galaxy clusters. 
Recent advancements in discrete dynamical modelling like Jeans dispersion-kurtosis, incorporating high-velocity moments \citep{Napolitano2014}, and chemo-dynamical modelling methods \citep{Zhu_nov2016} allow the unification of several physical properties of discrete tracers at once, like their positions, velocities, as well as colours and metallicities. This has brought a significant improvement and produced better constraints on the mass modelling and orbital anisotropy of the tracers \citep{Watkins2013, Zhu2016, Zhu_nov2016}.

Recently, \cite{Chao2020} used GCs kinematics to study the mass distribution and kinematics of the giant elliptical galaxy M87 in the centre of the Virgo cluster, based on 894 discrete tracers.
The M87 GC system (GCS) and the core of the Virgo cluster are a well-explored environment in this respect. 

A very interesting and dynamically evolving environment is the Fornax galaxy cluster, the most massive galaxy overdensity within 20 Mpc after the Virgo cluster. It is an ideal target to study the effect of the environment on the structure and assembly of galaxies of any type in detail \citep{Iodice2016, Iodice2019}.  The earliest approach to dynamically model the Fornax central galaxy NGC\,1399, was done by \cite{Kissler1999}, \cite{Saglia2000} and \cite{Napolitano2002}. Later, major and crucial work was presented by \cite{Schuberth2010}, where they used 700 GCs within 80 kpc of the Fornax cluster as dynamical tracers to put constraints on the properties of the central DM halo. 

In the last decade, the GC system of the Fornax cluster has been examined in great detail photometrically. Various imaging surveys like the ACS Fornax Cluster Survey (ACSFCS) with the Hubble Space Telescope \citetext{\citealp{Jordan2007}, also see \citealt{Puzia2014}}, the Next Generation Fornax Survey (NGFS) \citep{Munoz2015}, and the Fornax Deep Survey (FDS) with the VLT Survey Telescope (VST) \citep{Iodice2016} have added a wealth of information about galaxies and GCs in the Fornax cluster. Photometric studies from \cite{Abrusco2016}, and \cite{Iodice2019} have revealed that despite the regular appearance of the Fornax cluster, its assembly is still ongoing, as evidenced by the presence of stellar and GC tidal streams.
Most recently, \cite{Cantiello2020} have produced the largest photometric catalogue of compact and slightly extended objects in the Fornax cluster, over an area of $\sim$27 square degrees.

Due to its proximity, Fornax provides a unique opportunity to map its complex kinematics from the core out to at least $\sim$350 kpc using GCs as kinematic tracers.
Using the radial velocities of GCs and PNe, the Fornax Cluster VLT Spectroscopic Survey (FVSS) has
obtained an extended velocity dispersion profile of the central galaxy out to 200 kpc \citep{Pota2018, Spiniello2018}.
This has allowed to connect the large-scale kinematics of the major galaxies to the small scale stellar halo kinematics of the central galaxy NGC\,1399.

A crucial missing information to comprehend the complete mass assembly of Fornax is to understand the origin and kinematical behaviour of its intra-cluster population and of the disturbed outer halos of interacting cluster galaxies. Several studies have shown that ignoring the presence of substructures, which are generated by accretion and merger events, impact the dynamical mass estimates of clusters, leading to erroneous cosmological inferences \citep{Old2017, Tucker2020}. For example, studying the kinematics of stellar populations in the core of the Hydra\,I cluster, \cite{Hilker2018} reported that small scale variations in the kinematics due to substructures can produce a notable change in the mass-modelling, leading to an overestimation of DM halo mass in the core of that cluster. Therefore, the identification and proper kinematical understanding of dynamically cold substructures and outer halos of interacting galaxies in clusters are essential to understand how these structures formed, assembled and evolved, and have to be taken into account for the mass modelling.

Using a novel cold structure finder algorithm, \cite{Gatto2020} made a first attempt to search for cold kinematical substructures in Fornax based on the GC and PNe data of \cite{Pota2018} and \cite{Spiniello2018}. This has revealed the presence of at least a dozen of candidate streams in the combined kinematical information of PNe and GCs dataset of the FVSS (Napolitano et al. in preparation). These substructures can then be subtracted from the underlying discrete radial velocity field in the Fornax core for unbiased dynamical models.
Since the work of \cite{Schuberth2010} about ten years ago, a major dynamical study of the Fornax cluster is still missing, and so far no disturbed halo features of central cluster galaxies and no intra-cluster substructures were taken into account in a thorough dynamical model of the Fornax cluster core.

The low recovery fraction of the GC radial velocity measurements from the earlier FVSS results \citep{Pota2018} and the improvement of the VIMOS ESO pipeline motivated us to reanalyze the VLT/VIMOS data of the Fornax Cluster.
Here, in this work, we present the radial velocity catalogue of GCs over an area of more than two square degrees corresponding to 250\,kpc of galactocentric radius. In forthcoming papers of this series (Chaturvedi et al., in prep.), we will discuss the identification and properties of substructures and the mass-modelling of the Fornax cluster with the final goal to understand its mass-assembly and dark matter halo properties.

This paper is organized as follows: In section \ref{sec2} we describe the observations and data reduction. The radial velocity measurements are presented in section \ref{sec3}, and the results in section \ref{sec4}. In section \ref{sec5}, we discuss the results and present the photometric and spatial distribution of our GC catalogue. Section \ref{sec6} summarizes our results and presents the scope of future work. In Appendices \ref{polychoise} and \ref{gc_portfolio} we describe some tests performed for the radial velocity analysis and an object portfolio used for visual inspection. In Appendix \ref{gc_catalog} we show an excerpt of the VIMOS data GC catalogue from this study. The full catalogue is available online. Throughout the paper, we adopt a distance to NGC\,1399 of D$\sim$19.95\,Mpc \citep{Tonry2001} which corresponds to a scale of 5.8 kpc per arcmin.

%--------------------------------------------------------------------
\section{Observations and Data Reduction}\label{sec2}

This work examines the detection and kinematical characterization of GCs in the Fornax cluster core within one square degree.  We have reanalyzed Fornax cluster VLT/VIMOS spectroscopic data taken in 2014/2015 via the ESO program 094.B-0687 (PI: M. Capaccioli). For a detailed description of observations preparation and target selection, we refer the reader to \cite{Pota2018}. Here we briefly summarize the observations details and present the new data reduction.

\begin{figure}
    \includegraphics[width=0.95\linewidth]{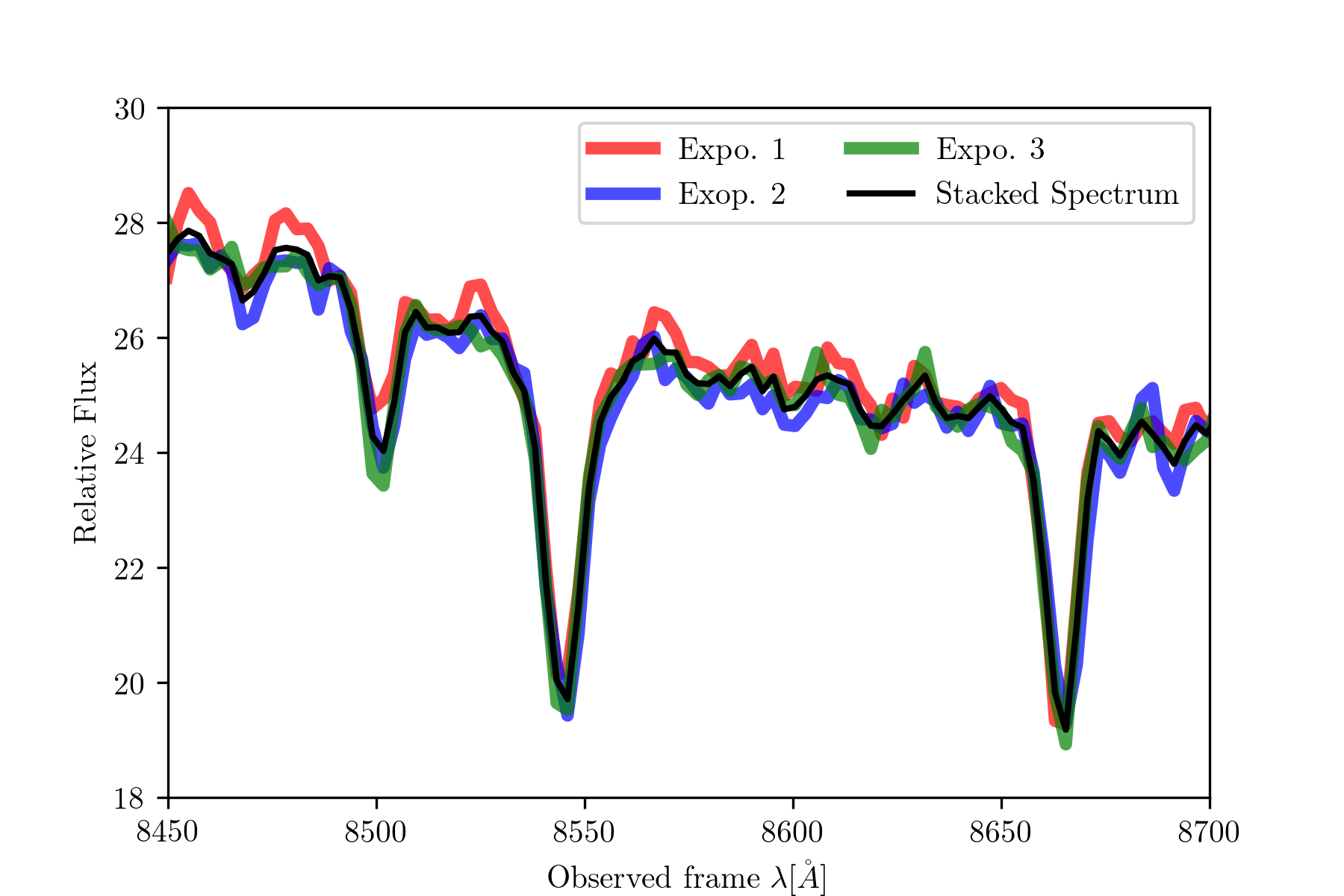}
    \caption{Individual and stacked reduced spectra for one example target. Red, green, and blue colours show the spectra for the three individual exposures. In black the final stacked spectrum is shown. No broadening is observed.}
    \label{fig:skyalign}
\end{figure}

\subsection{Photometry and globular cluster selection}
The Fornax Deep Survey \citep[FDS,][]{Iodice2016} and Next Generation Fornax Survey \citep[NGFS,][]{Munoz2015} formed the photometric data base to select GC candidates for the VIMOS/VLT spectroscopic survey. 
The FDS deep multiband ($u$, $g$, $r$ and $i$) imaging data from OmegaCAM cover an area of $\sim$30 square degrees out to the virial radius of the Fornax cluster. NGFS is an optical and near-infrared imaging survey and covers nearly the same area as the FDS survey. \\
GC candidates for spectroscopic observations were selected based on de-reddened $g$ and $i$ band magnitudes from the FDS and preliminary VISTA/VIRCAM photometry in the $K_{s}$ band from the NGFS. Additionally wide-field Washington photometry \citep{Dirsch2004, Bassino2006} was used to construct the $C-i$ vs $i-K_{s}$ diagram to select the bonafide GCs. Hubble Space Telescope/ACS photometry was used to find additional GCs in the central regions of NGC\,1399 \citep{Puzia2014}. Finally, with a magnitude restriction of $17.0<i<23.0$ mag, a total of 4340 unique spectroscopic targets were selected. This selection includes, on purpose, also some background galaxies and compact sources outside the selection criteria whenever there was space for additional VIMOS slits.

%--------------------------------------------------------------------
\subsection{Observations}

The spectroscopic observations for this study were carried out with the Visible Multi Object Spectrograph \citep[VIMOS,][]{LeFevre2003} at the VLT and were acquired in ESO Period 94 between October 2014 and January 2015. A total of 25 VIMOS pointings were defined to cover the central square degree of the Fornax cluster. Each pointing consists of 4 quadrants of 7$\times$8 arcmin$^{2}$. The MR grism was used together with filter GG475, which allows a multiplexing of two in spectral direction with a spectral coverage of 4750-10000$\AA$ at 12$\r{A}$ FWHM resolution. Three exposures of 30 minutes each were taken for each pointing.

%--------------------------------------------------------------------
\subsection{Data Reduction}

The data reduction was performed with the VIMOS pipeline version 3.3.0 incorporated in the ESO Reflex workflow \citep{Freudling2013}. The reduction follows the steps as described in \cite{Pota2018}. The dataset of each VIMOS pointing consisted of biases, flat fields, scientific spectral images, and arc lamp spectra. 
The older version of the VIMOS pipeline, used for the analysis performed in \cite{Pota2018}, did not correct for flexure induced wavelength shifts in multiple science exposure before their combination. This caused an incorrect absolute wavelength calibration and line broadening in the stacked spectra. \cite{Pota2018} manually corrected for this limitation, by applying the median wavelength shift of the second science exposure to the final stacked spectra. 
On the other hand, the improved pipeline version we used for our work takes care of wavelength shifts before stacking the individual spectra and prevents the line broadening effects. We confirmed this by reducing the individual exposures of a few VIMOS pointings and compared them with the final stacked reduced spectra. 

In Figure\,\ref{fig:skyalign} we show the CaT region of the reduced individual spectra and the stacked one for one example case. Red, green, and blue colours show the three single exposures. The black colour shows the final stacked spectrum. We repeated this test for several masks of different pointings, and no significant broadening was noticed. We checked this quantitatively by fitting a Gaussian to the CaT line at 8842 $\AA$ to the stacked and individual exposures of spectra with different signal-to-noise (S/N). The scatter among the mean positions of the CaT at 8842 $\AA$ line was found to be within the 10\% spaxel resolution limit of 2.58 $\AA$ for the used VIMOS grism. In order to obtain the final wavelength calibration, we provided our own skylines catalogue to calculate the residual shifts from the sky emission lines and corrected them.

\begin{figure*}[h]
\centering
\includegraphics[width=0.49\linewidth]{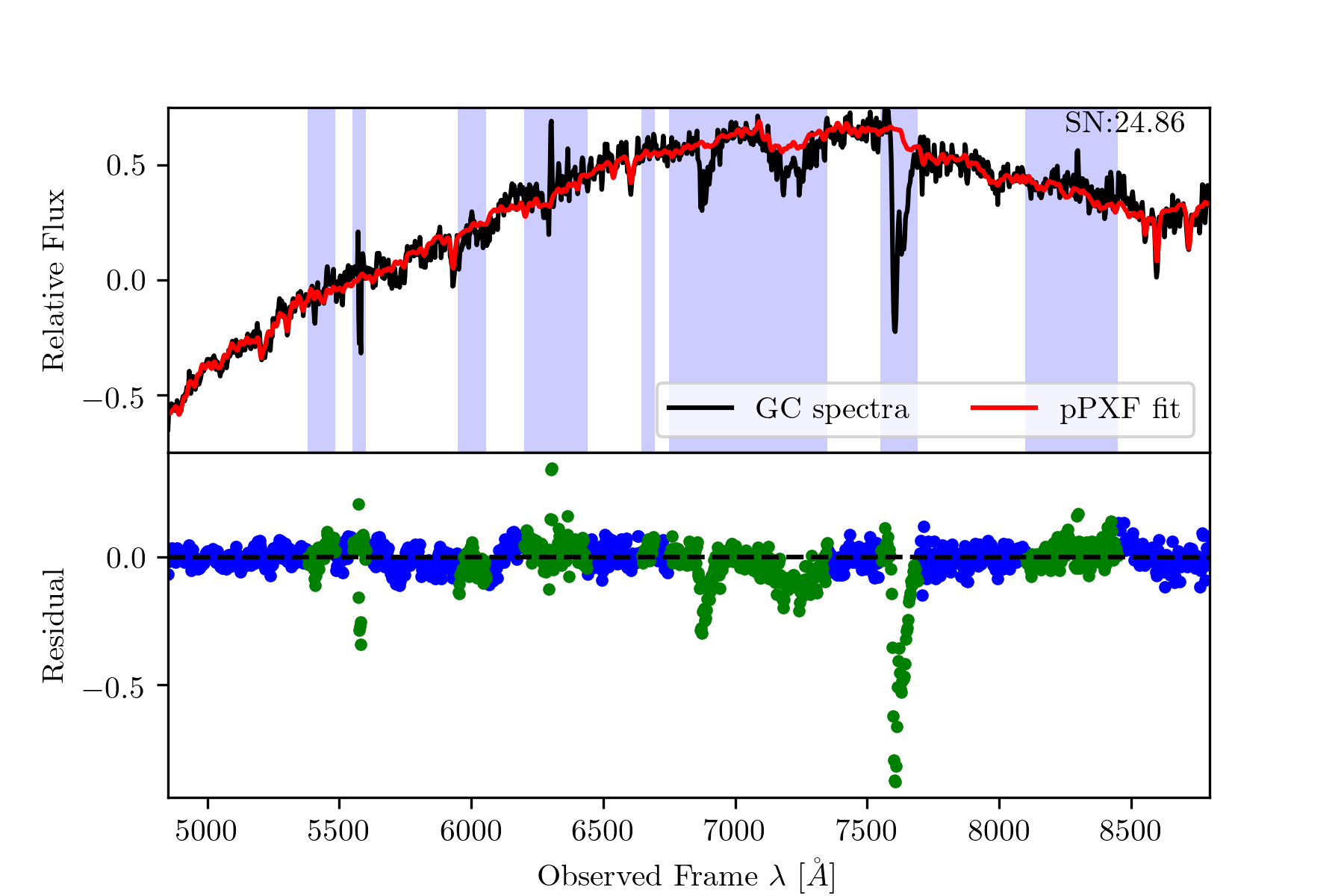}
\includegraphics[width=0.49\linewidth]{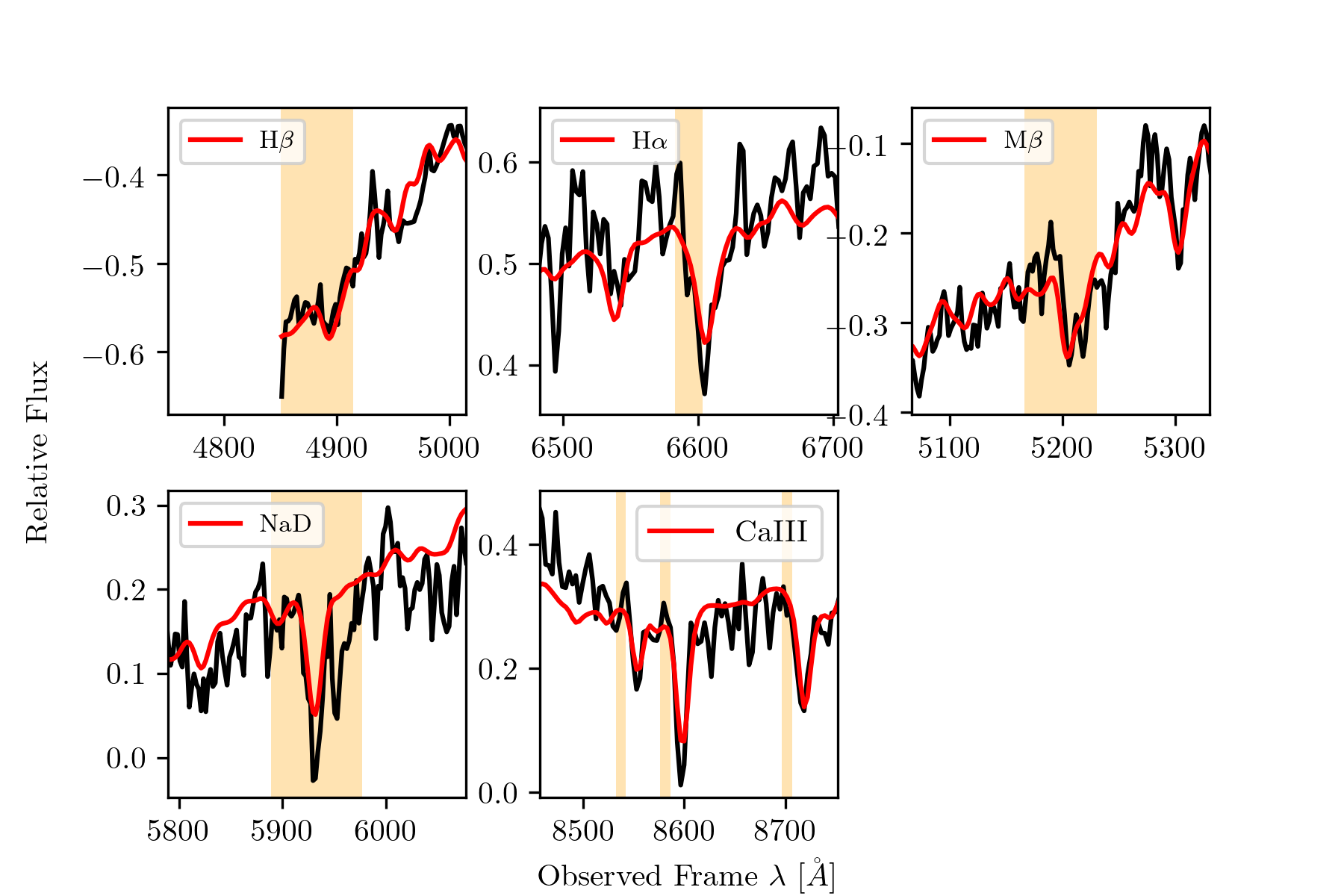}
\includegraphics[width=0.49\linewidth]{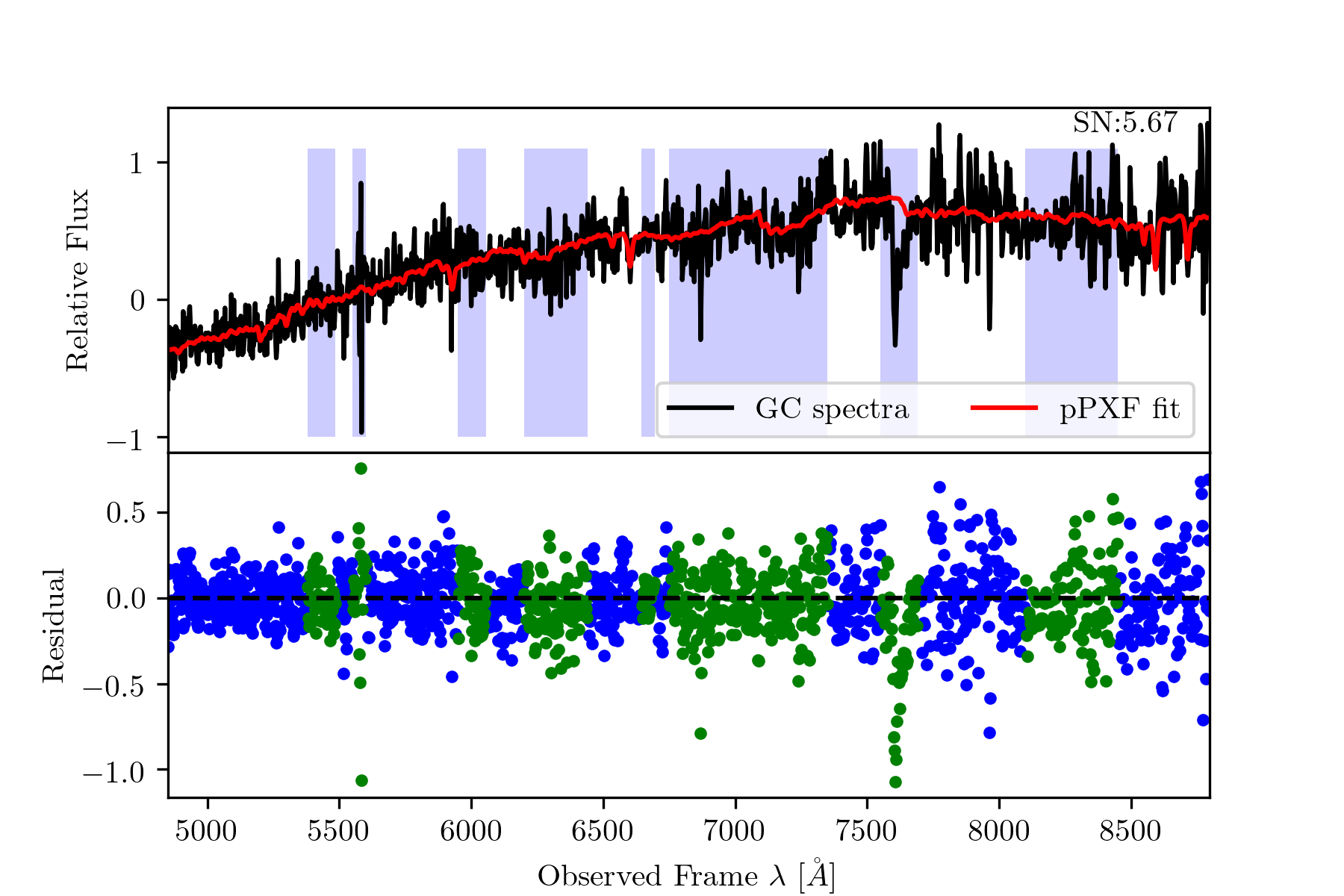}
\includegraphics[width=0.49\linewidth]{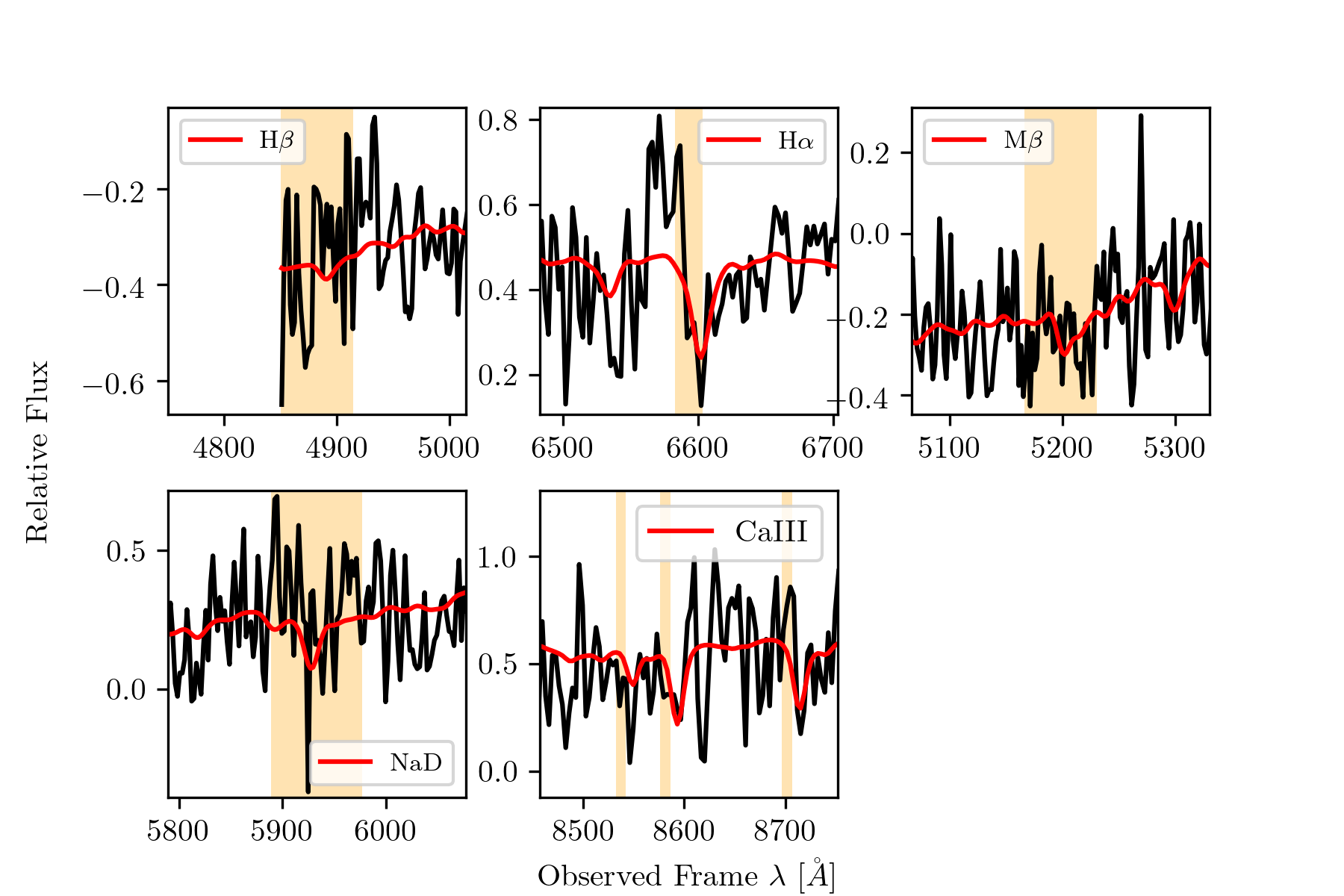}
\caption{Examples of pPXF fits to two GC spectra with different S/N ratios. In the top left panel, red and black colours show the pPXF fit and a GC spectrum with S/N$\sim$25. Masked regions are shown as blue bands. Blue and green dots in the lower sub-panel show the residuals for unmasked and masked regions, respectively. The top right panel shows a zoom-in view for the absorption features of H$\beta$, Mg$\beta$, NaD, H$\alpha$, and the CaT lines. Orange bands in the sub-panels show the expected position of absorption features at the Fornax cluster redshift. The two lower panels show the same, but for a spectrum with S/N $\sim$6.}
\label{fig:ppxffit}
\end{figure*}

%--------------------------------------------------------------------
\section{Analysis}\label{sec3}
In this section, we describe our analysis of the VIMOS data to obtain the line-of-sight (LOS) velocities from the spectra. We discuss our methodology for disentangling the GCs from background and foreground objects.

%--------------------------------------------------------------------
\subsection{Radial velocity measurements}\label{ppxf_los}
The radial velocity measurements of GCs were calculated using the python implemented penalized-PiXel fitting (pPXF) method of \cite{cappellari2004, cappellari2017}. pPXF is a full-spectrum fitting technique which generates a model spectrum by convolving a set of weighted stellar templates, to the parametric LOS velocity distribution (LSOVD), modelled as a Gauss-Hermite series. The intrinsic velocity dispersion of GCs (usually $\sim$20\kms) is well below the spectral resolution of the used VIMOS grism ($\sim$88\kms). Our initial test of deriving the radial velocities shows that we obtained a velocity dispersion always lower than 20\kms\ (i.e. in most cases pPXF gives a value of zero if the velocity dispersion is not resolved), which is an expected value for most of the GCs. We derive the mean velocity from pPXF and use the velocity dispersion value as a limiting criterion to select GCs.  

For the stellar templates, we used the single stellar population spectra from the extended medium resolution INT Library of Empirical spectra (E-MILES \citealt{Vazdekis2010, Vaz2016}) covering a wavelength range of 1680 to 50000 $\AA$. We preferred this stellar library, as it provides us flexibility in obtaining the stellar spectrum on a grid of ages ranging from 8 to 14 Gyr and metallicities in the range $-2.27<[M/H]<+0.04$ dex. We used an MW like double power law (bimodal) initial mass function with a mass slope of 1.30. With these settings, we obtain a set of 84 stellar templates from the E-MILES library at a spectral resolution of 2.51\,$\AA$. We convolved the stellar library with a Gaussian filter of standard deviation $\sigma=$12\,$\AA$ to bring it to the same resolution as the VIMOS spectra. 

For the spectral fitting, we use a wavelength region of 4800-8800 $\AA$. This wavelength region covers the major absorption line features, like H$\beta$ (4862 $\AA$), Mg$\beta$ (5176\,$\AA$), NaD (5895\,$\AA$), H$\alpha$ (6564\,$\AA$), and CaT lines (at 8498, 8548, 8662\,$\AA$). We masked several regions to avoid residual sky lines or telluric lines. 

pPXF requires starting values of the velocity moments parameter; in our case, its radial velocity and velocity dispersion. pPXF uses the given redshift of an object to make this initial guess (see section 2.3 \citealt{cappellari2017}). For the velocity dispersion, we chose a value of 0 km/s, which is expected for a GC (i.e. its internal velocity dispersion is not resolved). For the Fornax cluster redshift, the initial radial velocity was chosen to be $\sim$1375 \kms, close to the radial velocity of Fornax central galaxy NGC\,1399, which is 1425 \kms (taken from \citealt{Graham1998}). However, pPXF does not produce a meaningful fit with this guess in the case of foreground stars or background galaxies. Moreover, GC in the outer halo of NGC 1399 or in the intra-cluster regions of the Fornax cluster can have radial velocities different from the initial guess by about $\pm$500 \kms.  Therefore, it is crucial to test how the pPXF initial velocity guess can affect the resulting radial velocities, especially for GCs lying in the intra-cluster regions. To check this, we measured the radial velocities of different GCs belonging to the intra-cluster regions as a function of different initial velocity guesses, as shown in figure \ref{fig:vel_guess}. We found that the change in the resulting radial velocities is within 5\%, and this variation is within the measured velocity error. This shows that the initial guess of pPXF does not impact our radial velocity measurements. We present this test in detail in appendix \ref{polychoise}.

pPXF allows the use of additive and multiplicative Legendre polynomials to adjust the continuum shape in the spectral calibration. As a first guess, we have used a 3rd and 5th order for additive and multiplicative polynomials, respectively. However in cases, where we obtained a radial velocity consistent with the Fornax cluster but a velocity dispersion higher than 20\kms, we varied the polynomials such that we can obtain a velocity dispersion lower than 20\kms. We did a quantitative analysis to see how varying polynomials in pPXF can affect the derived mean velocities and dispersions. We did a test as follows: We defined a grid of additive and multiplicative polynomials in the range from 0 to 6. For each pair of polynomials, we derive the mean velocity and dispersion. First, we make sure that the velocity dispersion is lower than 20\kms\ and the variation in the mean velocity is not more than 5\% for different pairs of polynomials. Based on these two conditions, we select the most suitable value of these polynomials. We also checked the effect of higher order polynomials on the derived radial velocity as well as not using multiplicative polynomials. We found that in all cases there are only subtle variations in the derived radial velocities; they are all consistent within 5\%. We present more details on these tests of varying the order of polynomials in appendix \ref{polychoise}.

Figure \ref{fig:ppxffit} shows examples of pPXF fits for two GC spectra with different S/N ratios. Masked regions are shown in the upper left panels as blue bands, and residuals from the masked regions are plotted in green in the panels below. The right panels show zoomed in views of individual absorption line features of the GC spectra. Uncertainties on the mean velocity were estimated through Monte-Carlo (MC) realizations of the GC spectrum. For each pPXF model fit, we generated 100 realizations of spectra by adding Gaussian noise equivalent to the root mean square (RMS) of the residuals of the best fit. pPXF also returns the weights of template stars used for obtaining the best fit. To save computational time, we used only template stars with non-zero weights (around $\sim$7-10) for performing the MC realizations.
 
\begin{figure}
\centering
\includegraphics[width=1.0\linewidth]{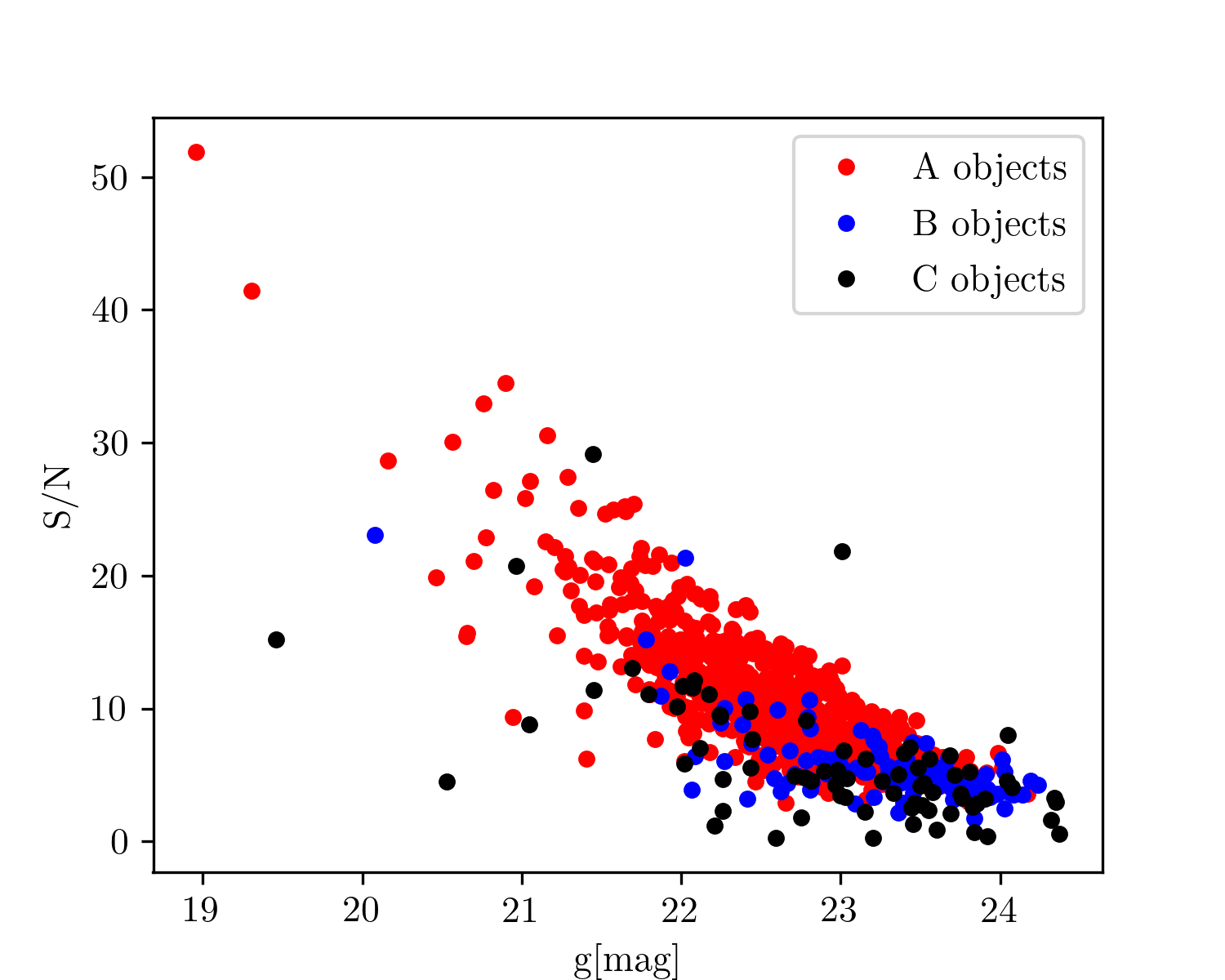}
\caption{S/N versus g-magnitude for the three classes of objects, as shown in the legend.}
\label{fig:sn_gmag}
\end{figure}

\subsection{Selecting Fornax Cluster Members}

Our VIMOS dataset is contaminated by foreground stars and background galaxies. To distinguish Fornax cluster GCs from the contaminants, we used the expected radial velocity range of $450<v<2500$\kms\ from \cite{Schuberth2010} for objects belonging to the Fornax cluster. 

We developed a two-step test to separate the GCs from the background galaxies and stars: 
First, we checked for the presence of emission lines in all spectra. In the case of multiple strong emission lines in a spectrum, we classify that object as a background galaxy. Second, the remaining spectra are fitted with an initial velocity guess of zero, and if pPXF returns a velocity lower than 450\kms, we classify the object as a star. All the remaining spectra are fitted as mentioned in section \ref{ppxf_los}. In this way we reject most of the contaminants at the very beginning, before deriving final radial velocity measurements.

To select the final bonafide GCs, we visually inspected the pPXF results for all remaining spectra. For that, we created a portfolio of each object with its 1D spectrum, pPXF fit with zoomed in views of major absorption features like H$\beta$, Mg$\beta$, NaD, H$\alpha$, CaT lines and a 2D image of the source, with attributes like S/N and radial velocity. One example of such a portfolio is shown in appendix \ref{gc_portfolio}. Based on these portfolios, we further classified the objects into three quality classes: Class A objects, where we can clearly see all the above-mentioned absorption features. Class B objects, where we see CaT and H$\alpha$ absorption features, but MgB and H$\beta$ are rarely recognizable. Class C objects, where we get Fornax cluster radial velocities, but hardly any absorption features are visible, although their colours are consistent with being GCs. Figure \ref{fig:sn_gmag} shows the S/N versus g-magnitude plot for A-, B- and C-class objects in red, blue and black, respectively. As expected, almost all the class A and B objects have S/N$>$3, whereas class C objects are mostly fainter and, on average, have a lower S/N ratio. Few C class objects have S/N$>$10, however their absorption features were contaminated by sky lines. We only consider class A and B objects for our kinematic analysis, but we include the class C objects in our catalogue. As the last step, we applied the heliocentric correction to all the bonafide selected GCs radial velocities, based on the header information of their observation date.

%--------------------------------------------------------------------
\section{Results}\label{sec4}
A total of 4574 slits were defined in the 25 VIMOS pointings. Around 2400 of them were allocated to GC candidates and compact objects, $\sim$800 slits to background galaxies and $\sim$1000 slits to stars \citep{Pota2018}. In our analysis, the Esoreflex pipeline extracted 5131 spectra from the VIMOS data (some slits contained more than one object), and our radial velocity analysis resulted in detecting around 920 possible Fornax cluster GCs. For the remaining spectra, around 1000 are classified as background galaxies, and approximately 700 objects revealed velocities of foreground stars. About 2500 spectra were of poor spectral quality, having either extremely low S/N, or being affected by strong residual telluric and skyline features.

We analyzed in detail the possible GC spectra. After visual inspection of the pPXF fits (section \ref{sec3}), we classified 839 spectra as class A and B objects, and 77 were classified as class C objects. After accounting for duplicate objects, we obtained 777 unique GC radial velocity measurements for class A and B objects. 

In appendix \ref{gc_catalog} we present an excerpt of our VIMOS GCs catalogue, including all A-, B- and C-class objects. The catalogue contains, despite the radial velocity, its error and S/N of the spectrum, also the photometric information in the $u$, $g$, $r$ and $i$ bands from the FDS \citep{Cantiello2020}. The full catalogue is available online.
In the following subsections, we present our results that are based on the new GC catalogue. 

\begin{figure}
\centering
\includegraphics[width=1.10\linewidth]{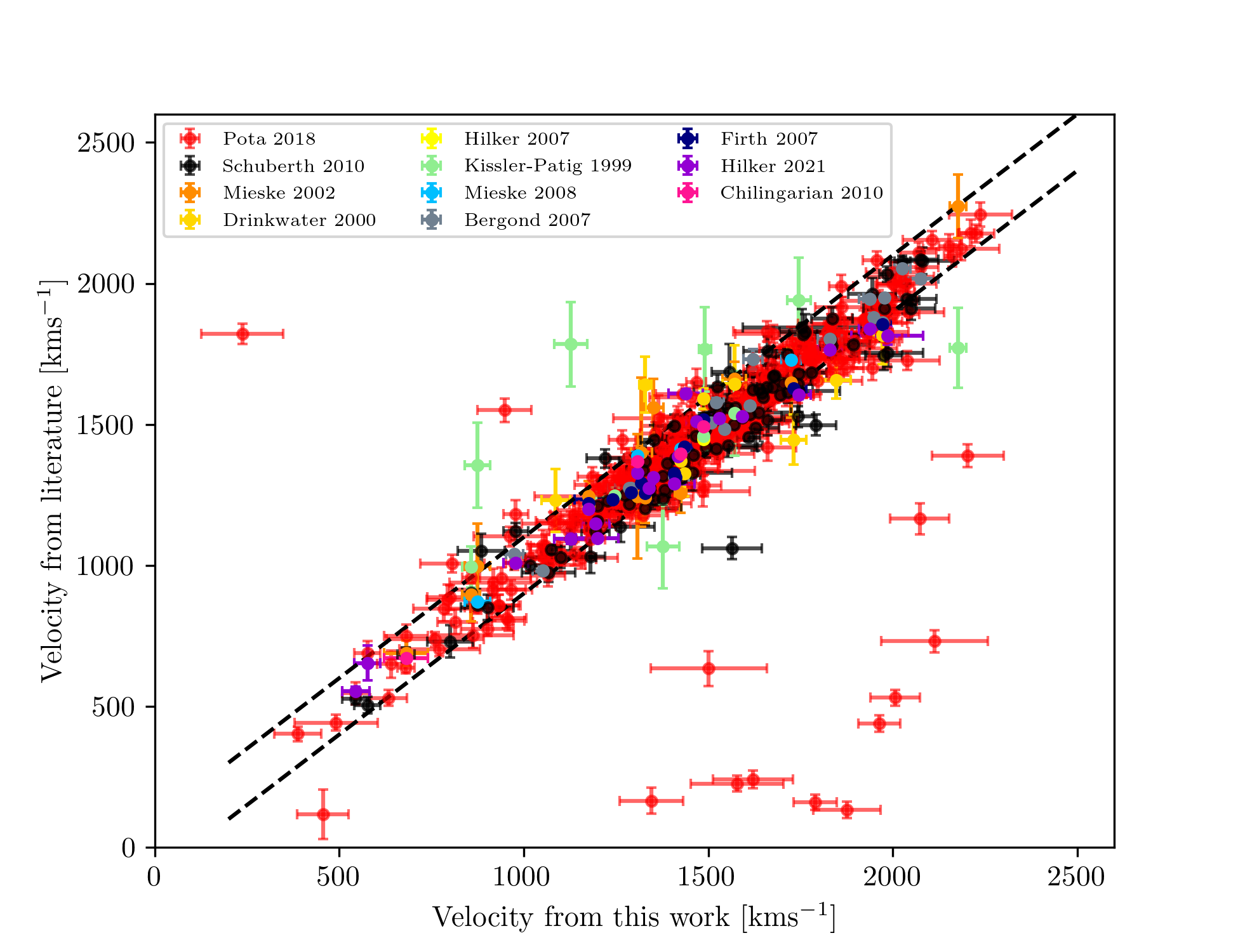}
\caption{Radial velocity comparison with previous measurements. The two dashed lines are drawn at $\pm$100\kms.}
\label{fig:velall_match}
\end{figure}

\subsection{Duplicate measurements}
\begin{figure}
\centering
\includegraphics[width=1.05\linewidth]{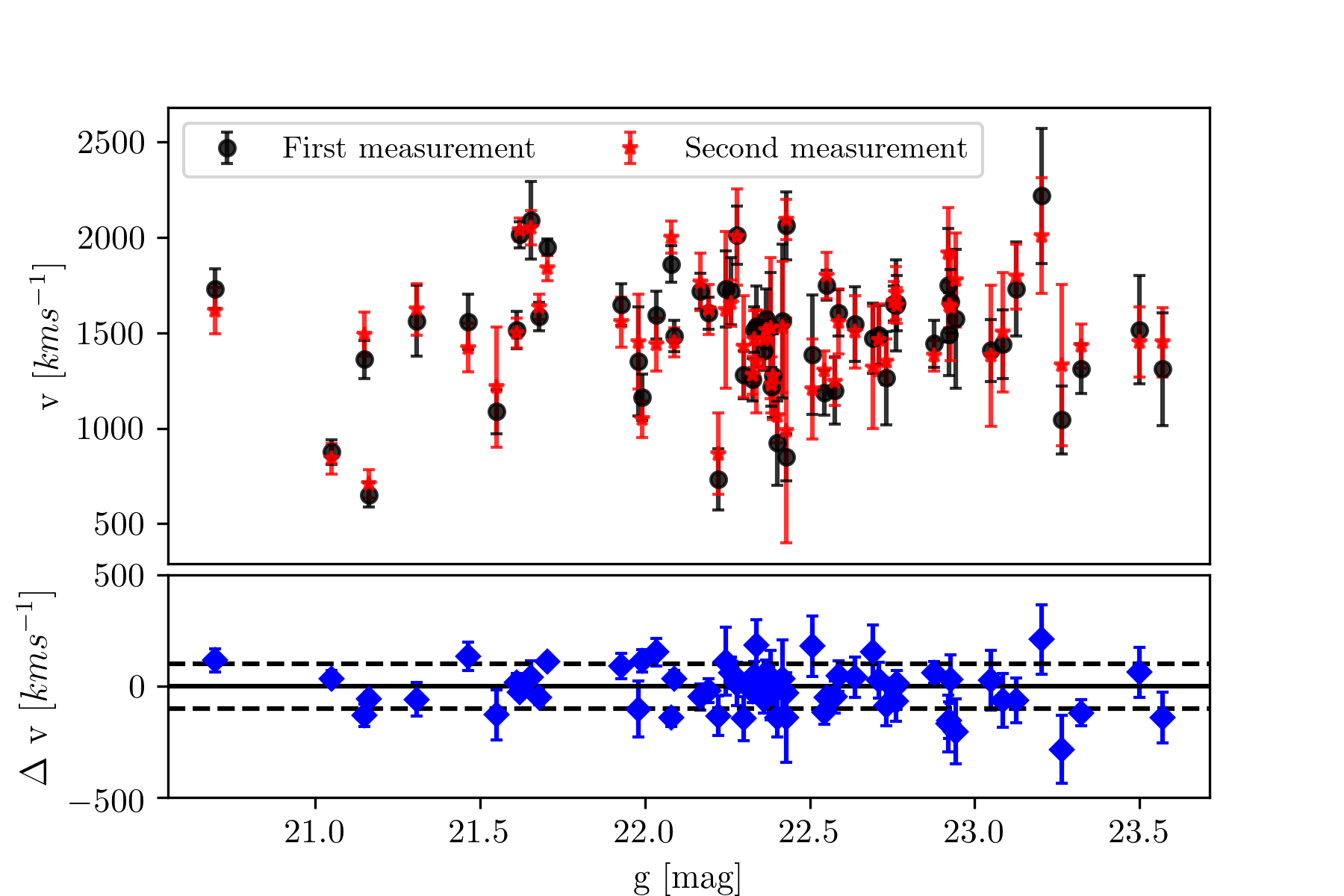}
\caption{Velocity comparison of duplicate measurements. Red stars and black dots show the radial velocity from two different measurements of the same object, as a function of $g$ magnitude. The bottom panel shows the velocity difference between the two measurements. The solid and dashed lines are drawn at $ \Delta$v = 0 and $\pm$ 100 $kms^{-1}$, respectively.}
\label{fig:dup_vel}
\end{figure}

We used the radial velocity measurements of the same objects observed in different VIMOS masks as a measure to check the robustness of the derived radial velocities and as an estimate for the errors. In figure \ref{fig:dup_vel}, we show the radial velocity measurements and their differences for 62 duplicate targets as a function of $g$ magnitude. 
The root mean square of the velocity difference is 104\kms, and the median offset is -11\kms. In case an object has a velocity difference of more than 3$\sigma$ of the corresponding uncertainty on the individual measurements, we take the velocity of the higher S/N spectrum; otherwise, we take the error weighted average velocity of two spectra. 

\begin{figure*}
\centering
\includegraphics[width=0.49\linewidth]{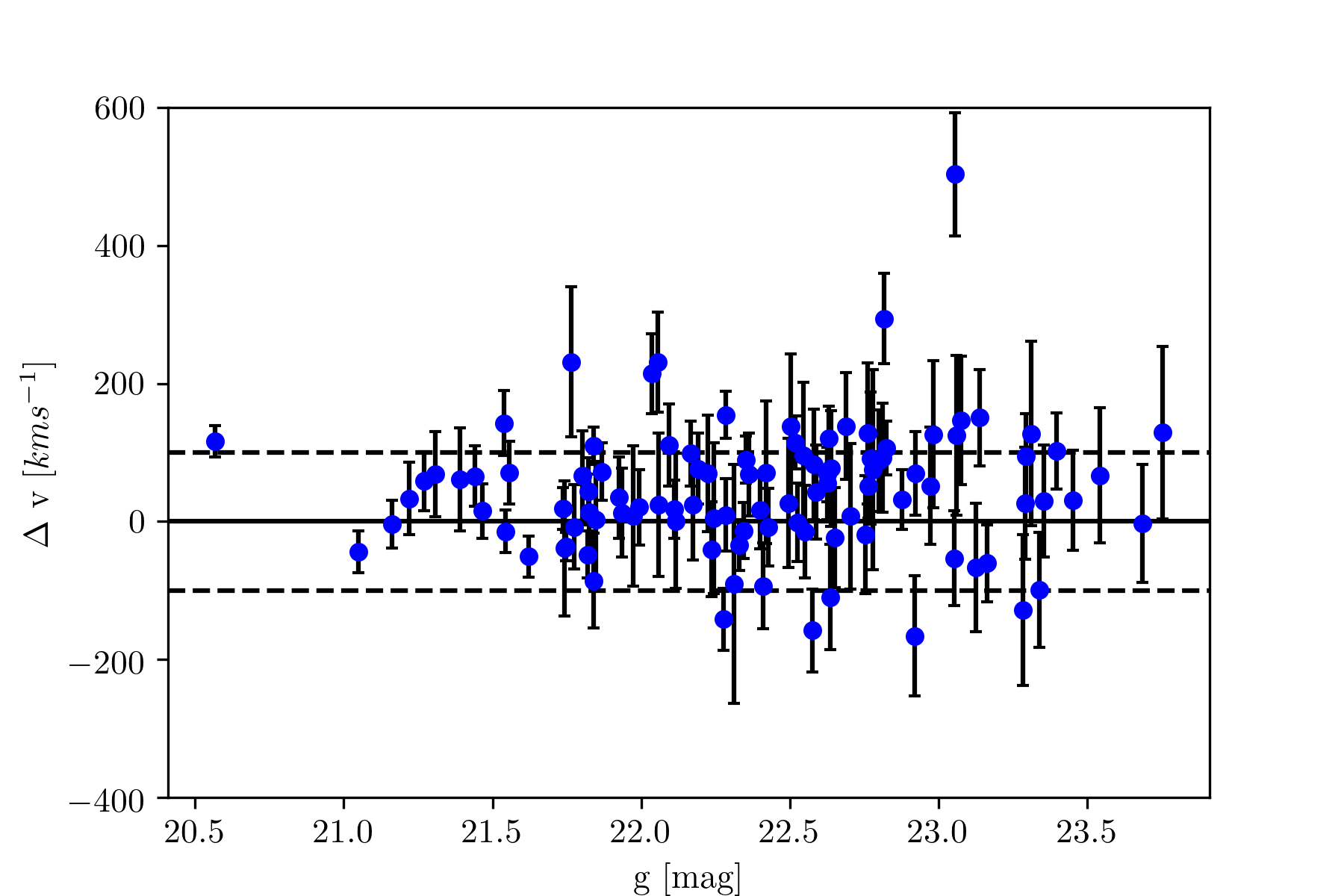}
\includegraphics[width=0.49\linewidth]{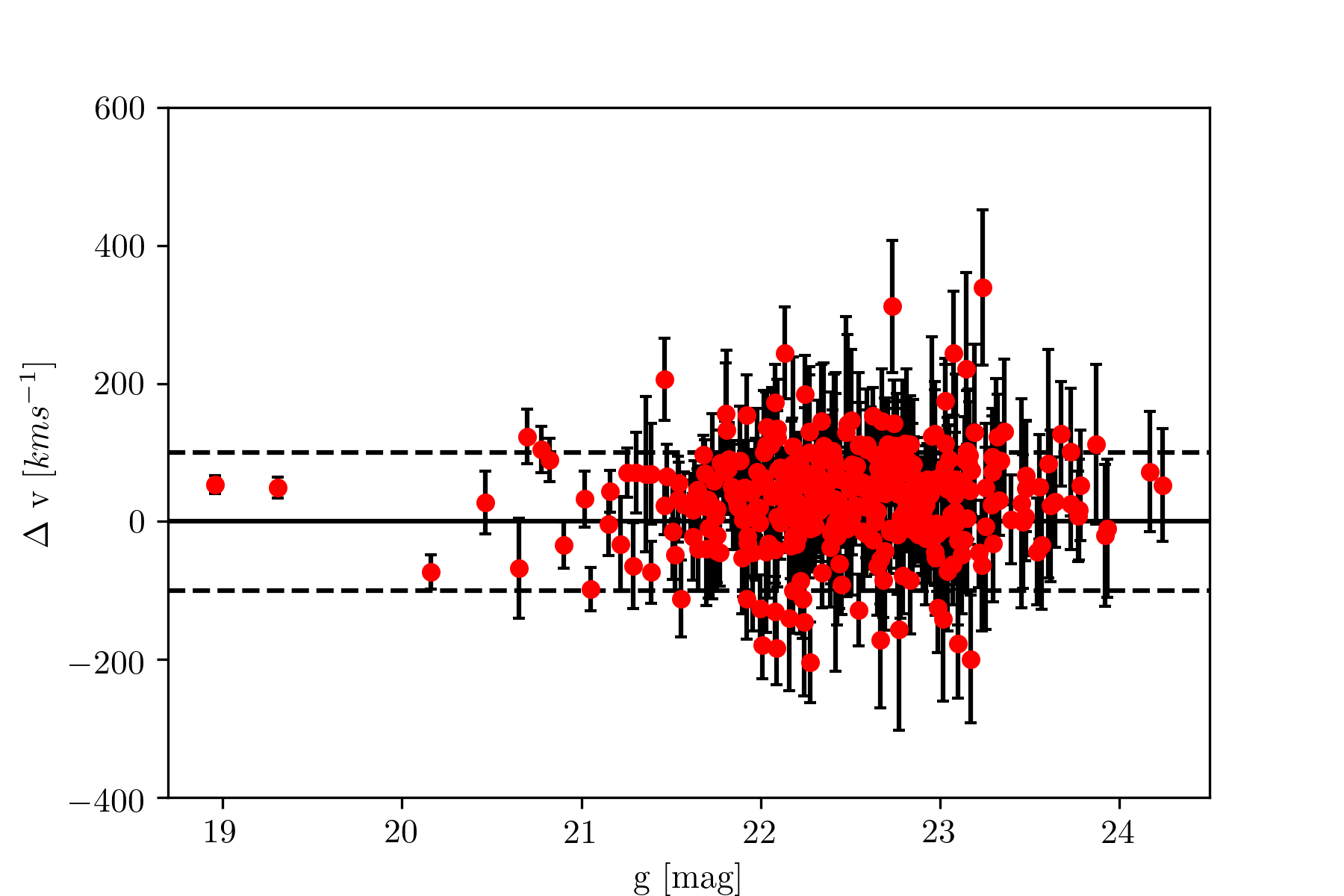}
\caption{Velocity measurement comparison with the GC sample of \cite{Schuberth2010} (left panel, blue dots) and that of \cite{Pota2018} (right panel, red dots) as function of $g$ magnitude. The solid and dashed lines are drawn at $ \Delta v$ = 0 and $\pm$ 100 $kms^{-1}$, respectively.}
\label{fig:sbrt_pota_match}
\end{figure*}

\subsection{Comparison with previous measurements}

\begin{figure*}
\centering
\includegraphics[width=0.49\linewidth]{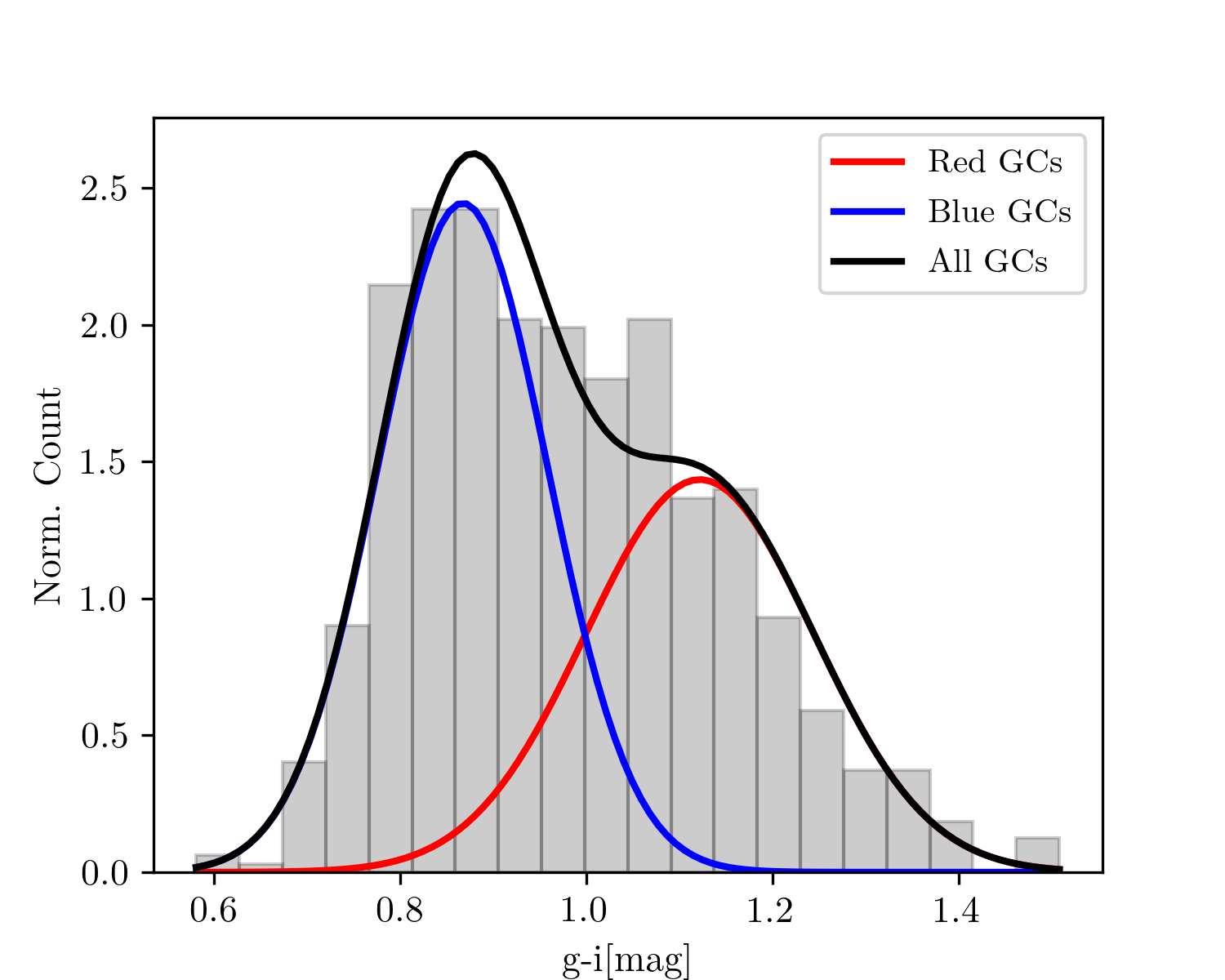}
\includegraphics[width=0.49\linewidth]{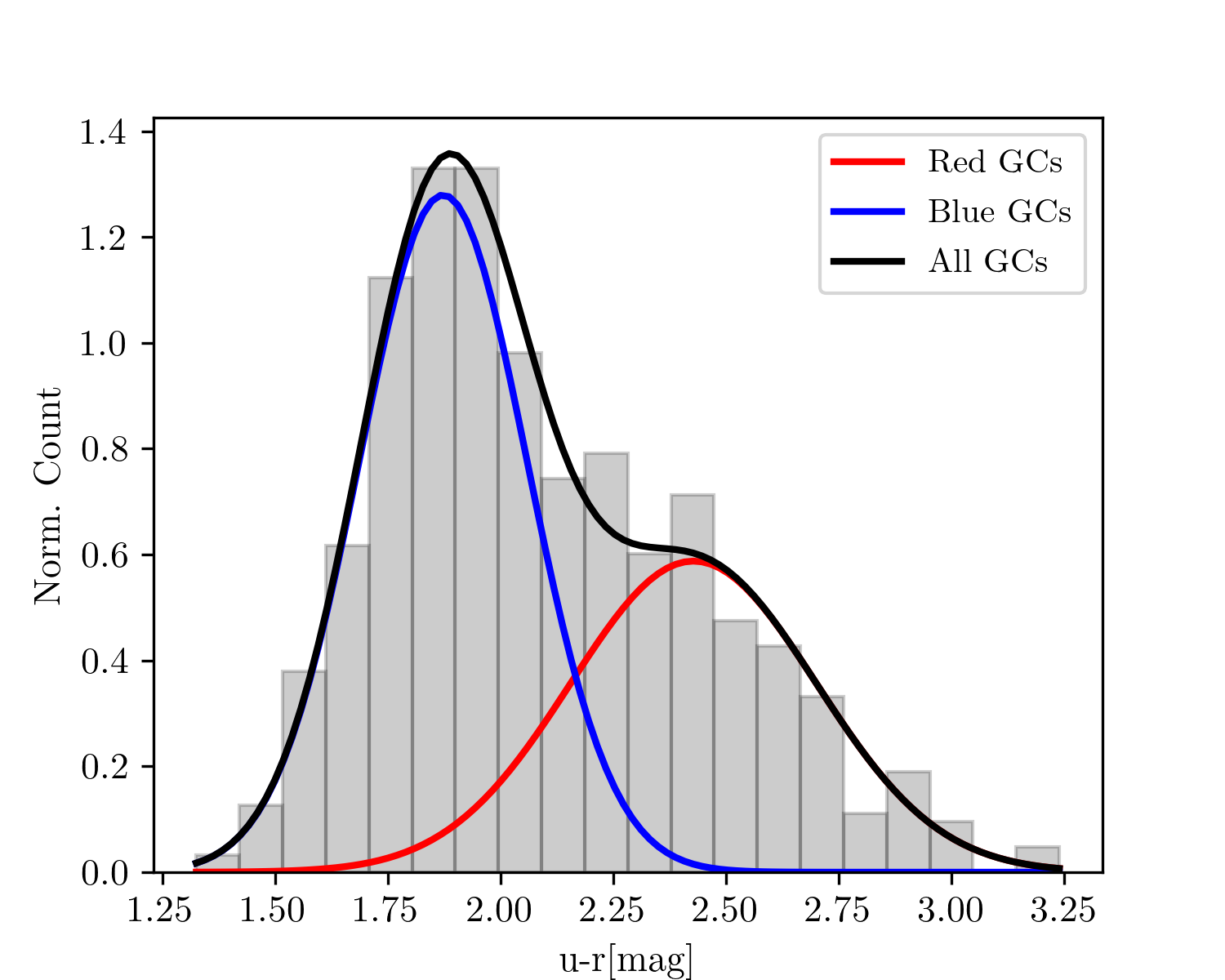}
\caption{Results of Gaussian Mixture Modeling. Left panel: Histogram and colour bi-modality of the GCs in $g-i$ colour distribution. Blue and red Gaussian curves are obtained from the GMM and represent the blue and red GC populations. Right panel: Same as left panel but for the $u-r$ colour distribution.}
\label{fig:red_blue_hist}
\end{figure*}

\begin{figure*}
\centering
\includegraphics[width=1.00\linewidth]{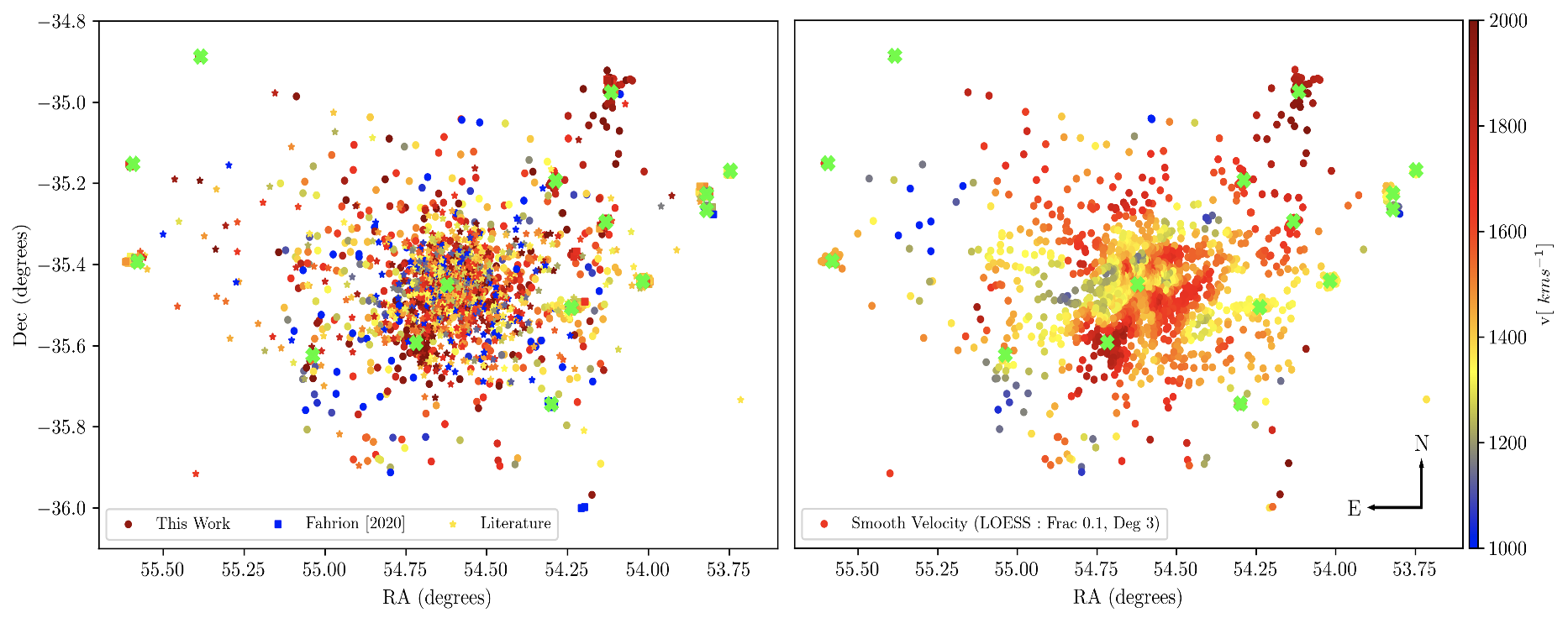}
\caption{Radial velocity map of GCs within 1.5 square degrees of the Fornax cluster. Major galaxies are shown with green crosses. Left panel: GCs from this work are shown as dots. Squares and stars show GCs from \cite{Katja2020} and previous literature measurements, respectively. Right panel: Smooth velocity map using the LOESS technique. The smoothing parameters are given in the legend of the plot.}
\label{fig:los_map}
\end{figure*}

Several past studies have probed the GC systems in the Fornax cluster. \cite{Schuberth2010} presented a catalogue of 700 GCs from observations with VLT-FORS2 and Gemini-GMOS. \cite{Bergond2007} measured the kinematics of 61 GCs in the intra-cluster space of the cluster based on FLAMES observations. Other studies \citep[like][]{Firth2007, Chilingarian2011} have targeted and analyzed the most massive compact stellar objects around NGC\,1399. These literature velocity measurements of GCs provide us another way to verify and check our derived radial velocity analysis's robustness.

Comparing our sample with \cite{Pota2018}, we obtained a match for 369 objects. Out of those, 22 objects were found to have a velocity difference of more than 3$\sigma$. Excluding the outliers, the RMS of the velocity difference is 72\kms\ and the median offset 32\kms. Here, the median offset is defined as the median of the velocity difference distribution between our GCs radial velocities minus the matched literature GCs velocities.

With the GC sample of \cite{Schuberth2010}, we obtained a match for 103 objects, of which only 5 were found to have a velocity difference of more than 3$\sigma$. Excluding these 5 outliers, we obtain an RMS of 80\kms\ and median offset of 43\kms. We visually inspected all the outliers in both samples and found that our fits to the spectra look very reliable and therefore neglect the previous measurements of \cite{Pota2018} and \cite{Schuberth2010} for the outliers. In figure \ref{fig:sbrt_pota_match} we show the velocity differences to both samples.

Finally, we compared our velocity catalogue with all other available literature studies. Table \ref{Tab1:vel_com} summarizes the number of matched objects with our velocity catalogue objects. Figure \ref{fig:velall_match} shows the velocity comparison between velocities measured in this work and from the previously available catalogues. 
We speculate that the measured mean velocity offsets between our and literature datasets might just be due to systematics in the zero points of the wavelength calibration, as it is common in multi-slit spectroscopy. All offsets are minor and within the overall velocity scatter, and therefore, we did not attempt to correct for them.

\begin{table*}
\caption[]{Matched number of objects from this study to previous studies, rms scatter and median offset of the comparison (our$-$previous work).}
\label{Tab1:vel_com}
\centering
\begin{tabular}{lccc} 
\hline
Previous Study           & Matches   & RMS (\kms)  & Median offset (\kms) \\  
        \hline
        \cite{Pota2018}          & 369 & 72 & 32 \\
        \cite{Schuberth2010}     & 104 & 80 & 43 \\
        \cite{Mieske2002}        & 13 & 102 & -38\\
        \cite{Drinkwater2000}    & 10 & 171 & 9\\
        \cite{Hilker2007}        & 1  & 38  & 38\\
        \cite{Kissler1999}       & 10 & 125 & -69\\
        \cite{Mieske2008}        & 5  & 38  & -3 \\
        \cite{Bergond2007}       & 18 & 59  & 21\\
        \cite{Firth2007}         & 11 & 68  & 33\\
        Hilker \& Puzia (priv.comm.) & 20 & 84 & 36 \\
        \cite{Chilingarian2011}  & 4  & 34 & 2\\
        \hline
  \end{tabular}
\end{table*}

\begin{figure}
\centering
\includegraphics[width=0.95\linewidth]{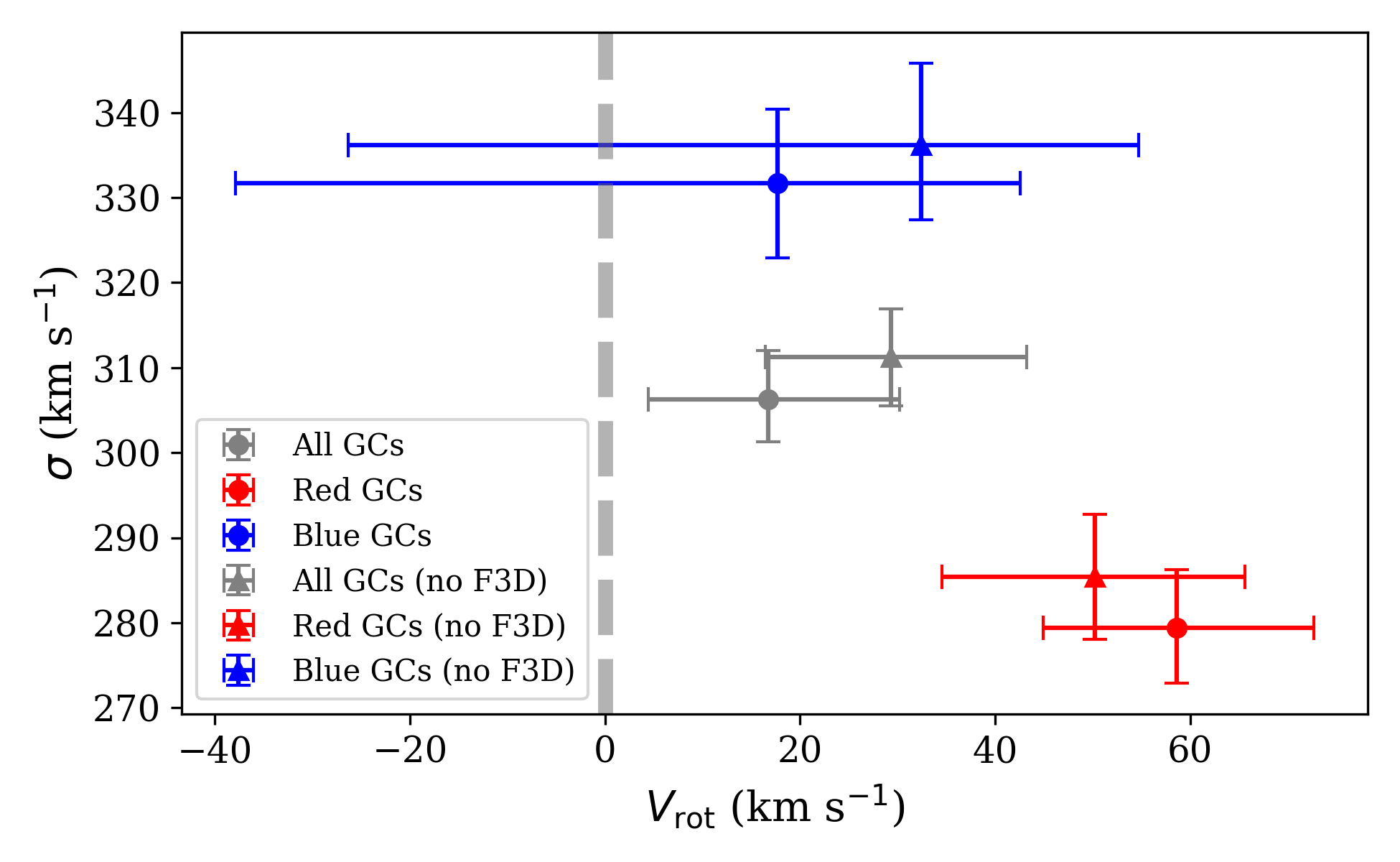}
\caption{Rotational velocity of all, red and blue GCs within 30 arcminutes. We modelled the full sample as well as a restricted sample, excluding the F3D GCs from \cite{Katja2020} as most F3D GCs are bound to their respective host galaxies.}
\label{fig:vrot}
\end{figure}

\subsection{Photometric properties}

\begin{figure*}
\centering
\includegraphics[width=0.50\linewidth]{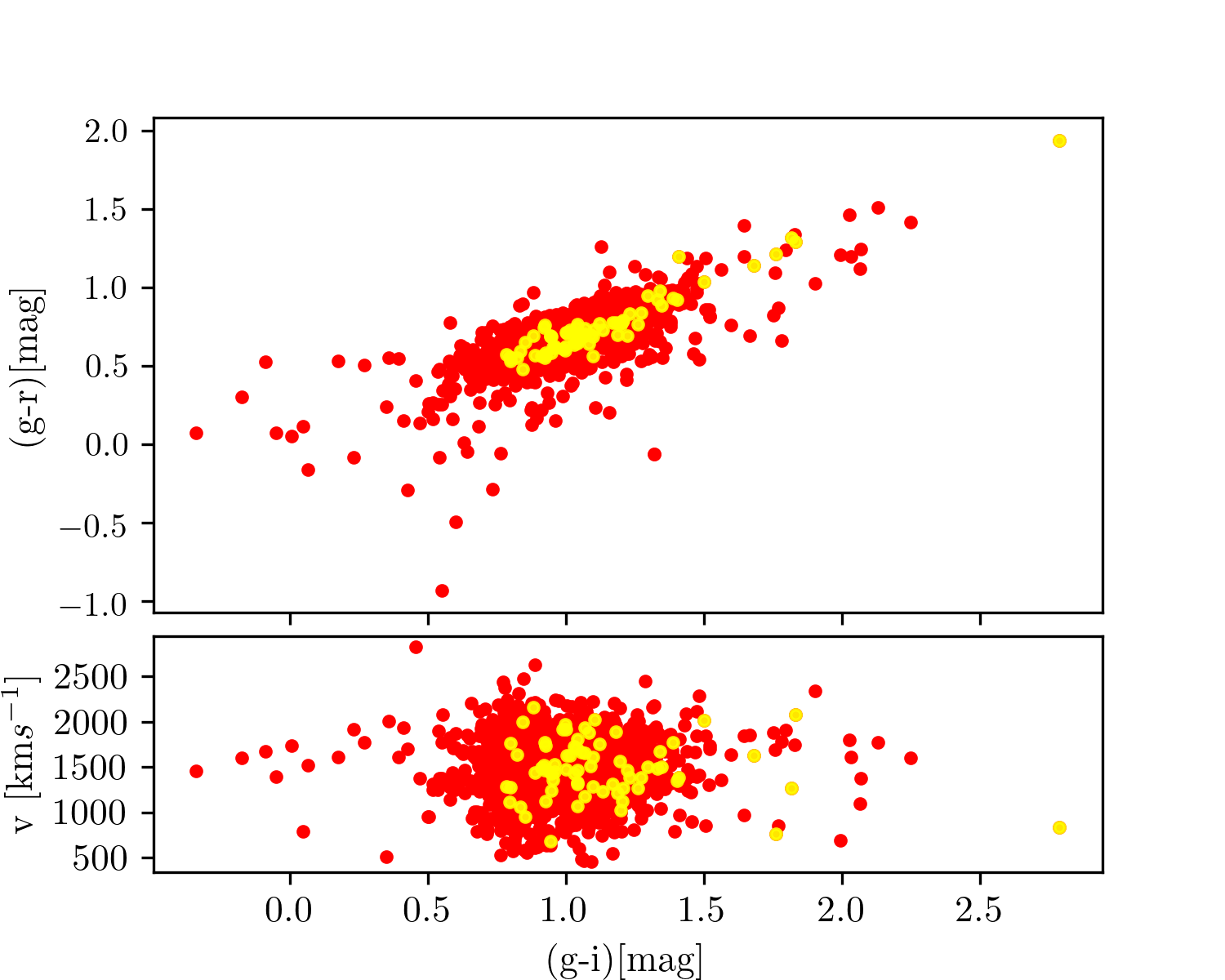}\hspace*{-2.0em}
\includegraphics[width=0.50\linewidth]{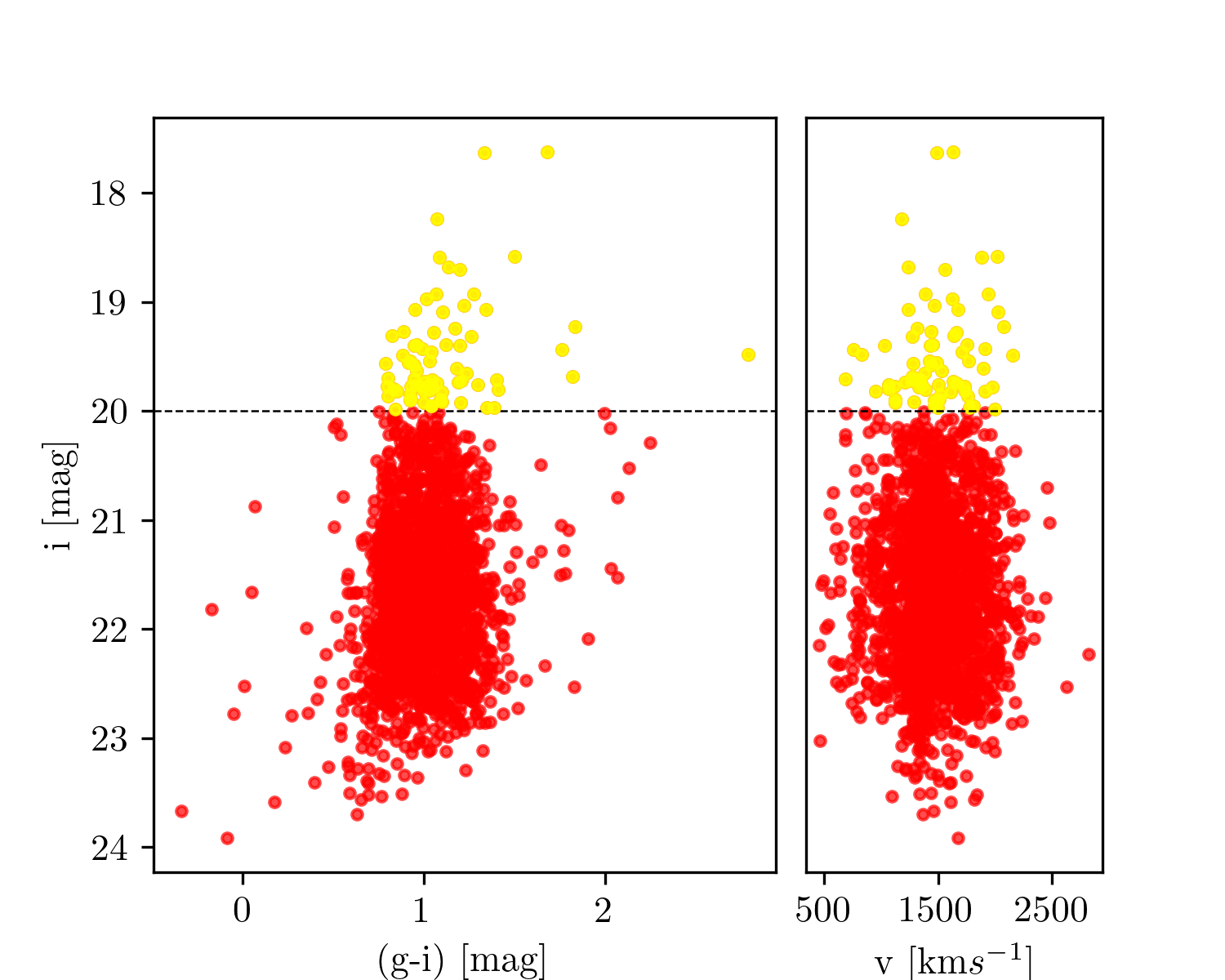}
\caption{Distribution of GCs (red dots) and UCDs (yellow dots) in the colour-colour and colour-magnitude diagram. Left panel: The distribution of GCs in the $(g-r)$ vs $(g-r)$ space. The lower subplot shows the radial velocity as a function of $(g-i)$ colour. Right panel: Same as right panel, but in $i$ mag vs. $(g-i)$ space.}
\label{fig:cm_cmd}
\end{figure*}

To get the photometric properties of our GC sample, we matched it with the photometric $ugri$ and $gri$ catalogues presented by \cite{Cantiello2020}. We get a photometric match for 700 and 770 objects with the $ugri$ and $gri$ catalogues, respectively. To separate our GC sample in blue and red GCs, we follow the procedure used by \cite{Angora2018} and \cite{Cantiello2020}, namely Gaussian Mixture Modelling (GMM) implemented through the python library $sklearn$ \citep{Pedregosa2012}. We fitted a bi-modal Gaussian distribution to the GC populations in the $u-r$ and $g-i$ colour-colour diagrams. Figure \ref{fig:red_blue_hist} shows the projected distributions of the bivariate Gaussian (and their components for blue and red GCs) on the $g-i$ and $u-r$ colour axes. 

A linear fit between the intersection of blue and red Gaussians for $g-i$ and $u-r$ is used to divide the GCs into the respective samples. Table \ref{Tab2:gmm_table} shows the results of our GMM. Out of 770 objects, our sample has 56\% blue and 44\% red GCs, as judged from the photometrically complete $gri$ sample.

\begin{table}
\caption[]{Bi-variate Gaussian parameters using GMM.}
\centering
\begin{tabular}{c|cc|cccc}

\hline
 & Blue & & Red & \\  
\hline
\hline

Parameter & $g-i$ & $u-r$ & $g-i$ & $u-r$\\  
\hline
    $\mu$   & 0.876 & 1.872 & 1.104 & 2.427 \\
    $\sigma$ & 0.009 & 0.034 & 0.016 & 0.074 \\
\hline
\end{tabular}
\label{Tab2:gmm_table}
\end{table}

\begin{figure*}
\centering
\includegraphics[width=0.52\linewidth]{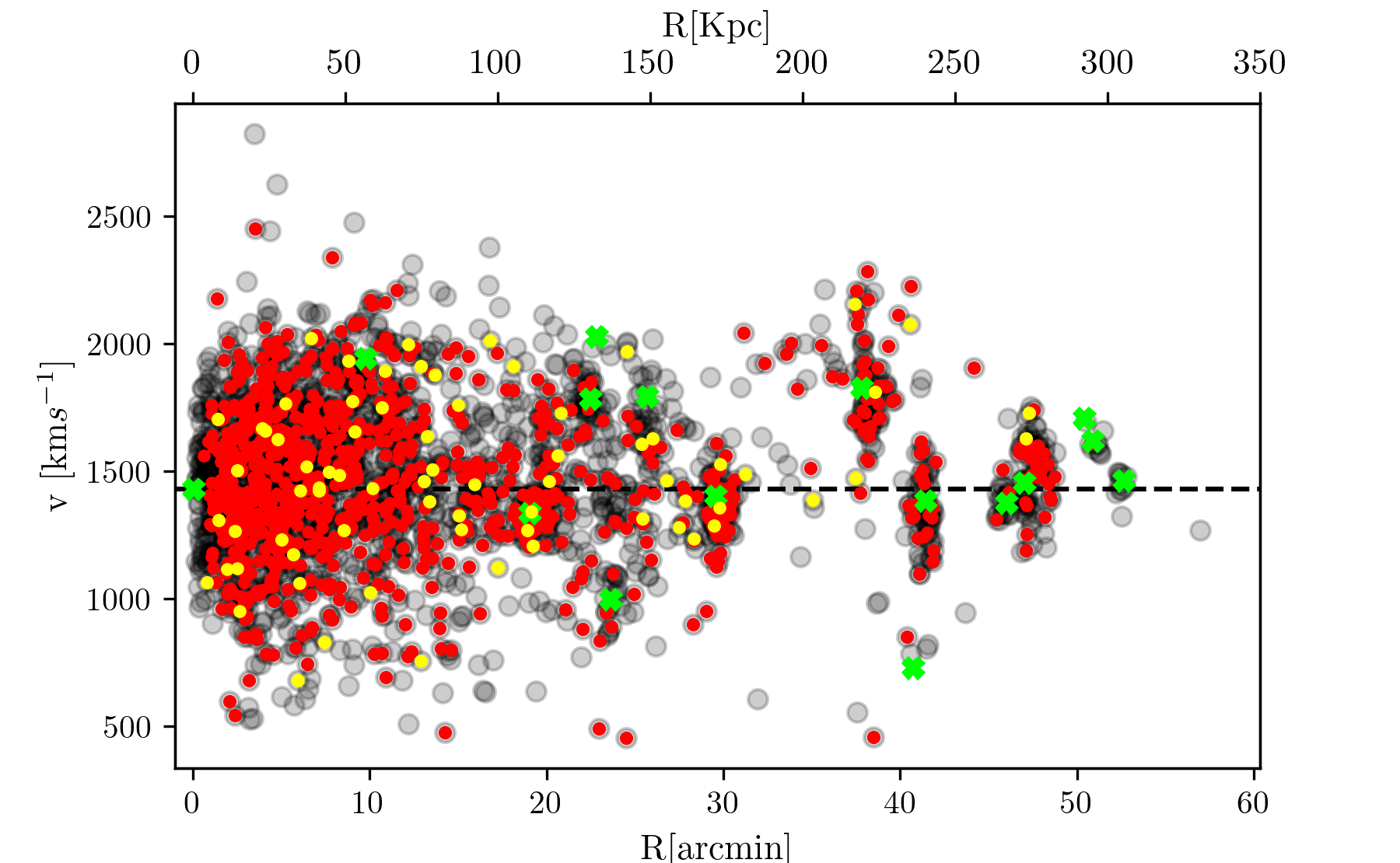}\hspace*{-1.0em}
\includegraphics[width=0.52\linewidth]{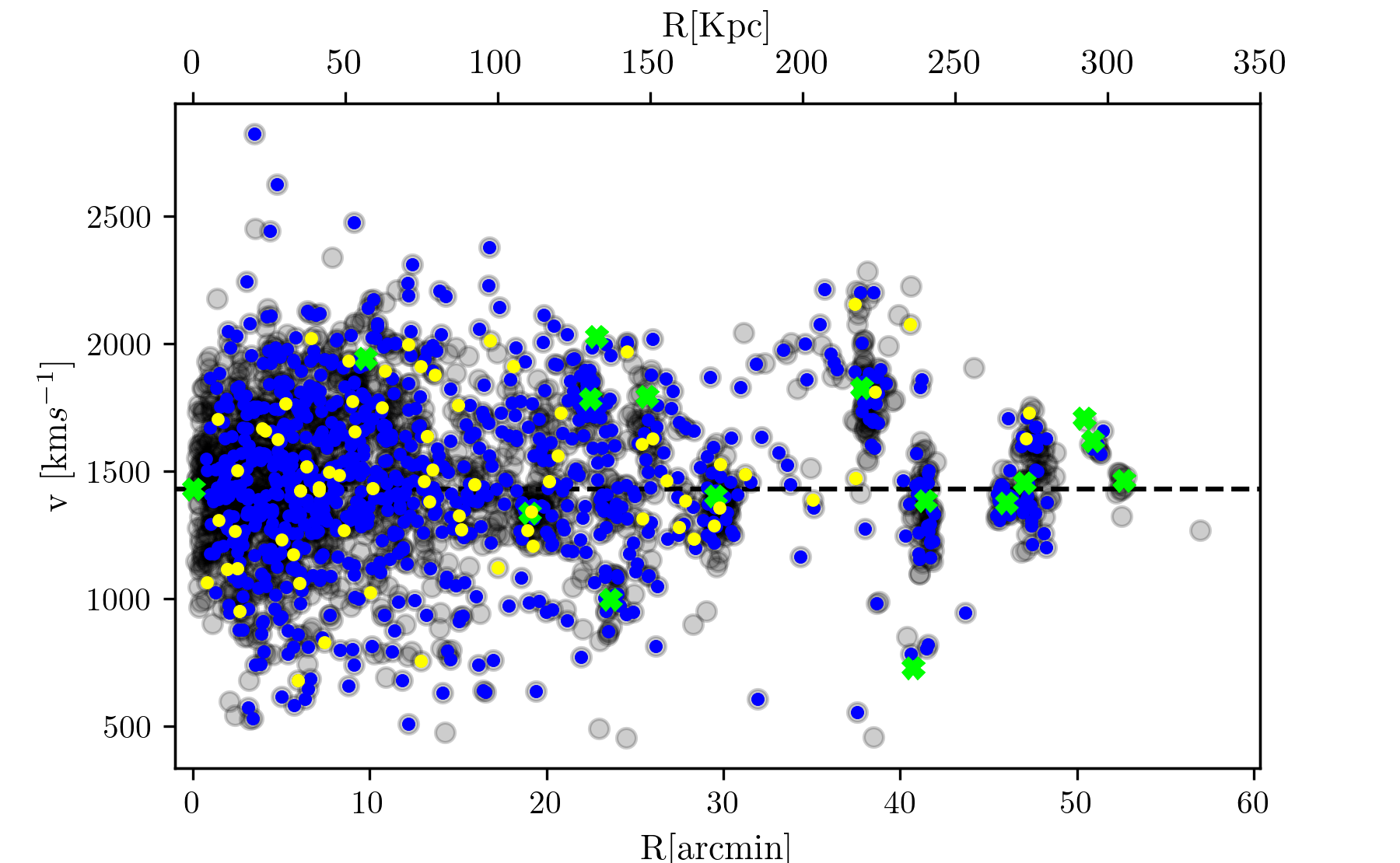}
\caption{Phase-space diagrams of radial velocity vs cluster-centric distance. In both panels the grey dots show the full sample, the limes crosses major galaxies within 300 kpc, and the dashed horizontal lines mark the systemic velocity of NGC\,1399. Left panel: Velocity distribution for red GCs. Right panel: The same for blue GCs. Yellow dots indicate UCDs, and the lime crosses mark the major galaxies of the Fornax cluster.}
\label{fig:rad_vel_dis}
\end{figure*}

\begin{figure*}[h]
\centering
\includegraphics[width=1.05\linewidth]{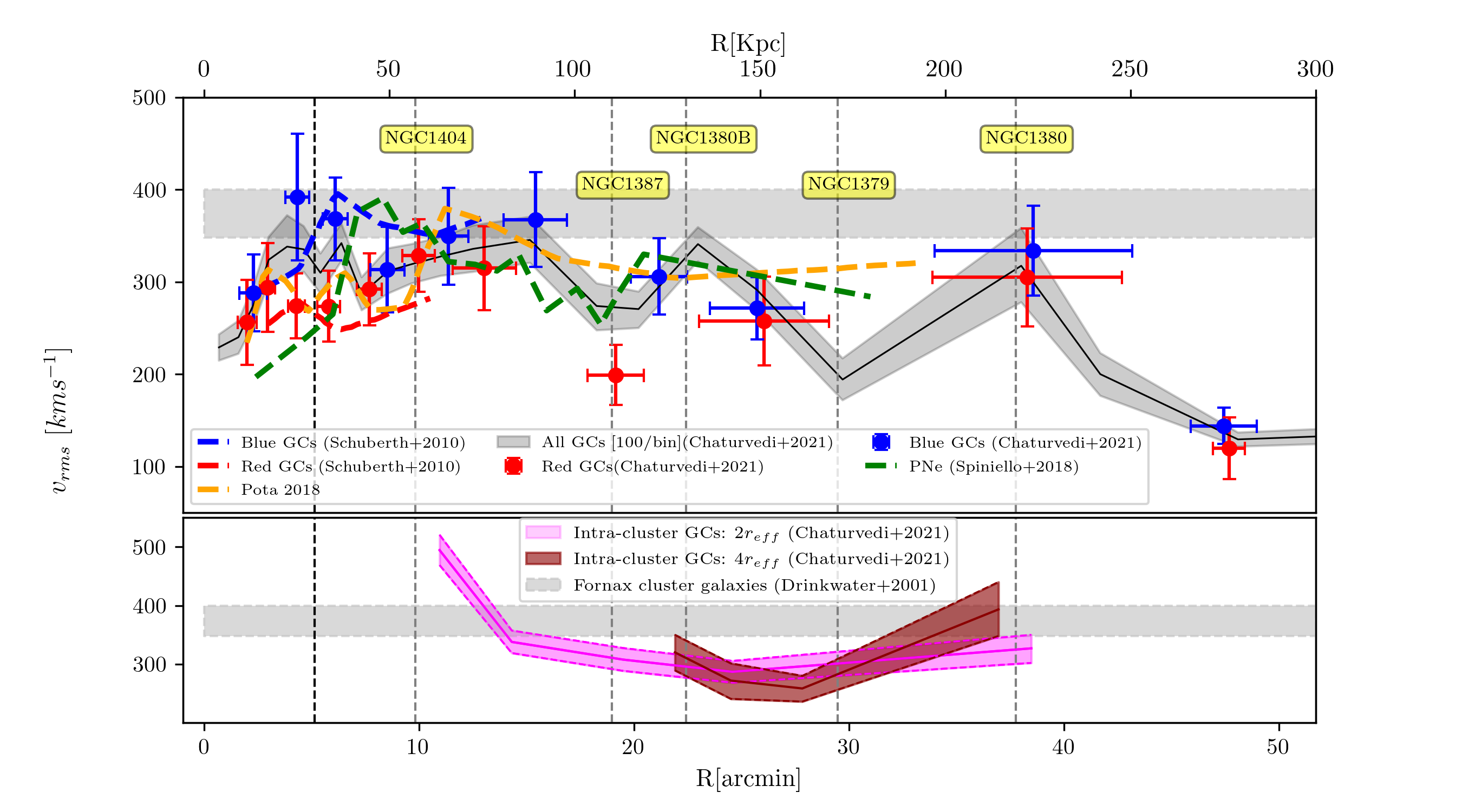}
\caption{Velocity dispersion profiles in the Fornax cluster core region as a function of projected distance from NGC\,1399. Upper panel: The black line denotes the dispersion profile of the complete sample of 2341 GCs. The grey band marks 1$\sigma$ uncertainty. Red and blue dots represent the values for the red and blue GCs, with 100 GCs per bin. The dashed blue and red lines represent the dispersion profiles for the GC analysis of \cite{Schuberth2010} data. The green and orange dashed lines show the PNe and GC dispersion profiles from \cite{Spiniello2018} and \cite{Pota2018}, respectively. The horizontal band denotes the velocity dispersion of the Fornax cluster galaxies \citep{Drinkwater2000}. The vertical black dashed line marks the effective radius of NGC\,1399 and the vertical grey dashed lines the projected distances of major galaxies (as labelled) from NGC\,1399. Lower panel: The dispersion profile of ICGC candidates with two different selections is shown. The pink line shows ICGCs and outer halo GCs selected further than 2$r_{\rm eff}$ away from major galaxies, and the light pink band represents the 1$\sigma$ uncertainty. The dark red line represents ICGCs that were selected outside 4$r_{\rm eff}$ around major galaxies (see text for details), and the lighter red band denotes its 1 $\sigma$ uncertainty.}
\label{fig:dis_pro}
\end{figure*}

\subsection{Radial Velocity Map}

Combining our radial velocity measurements with previous literature measurements and the recent catalogue presented by \cite{Katja2020}, brings the total number of confirmed GCs in the Fornax cluster to 2341 objects. The catalogue of \cite{Katja2020} is based on integral-field observations of the Fornax3D project (F3D, \citealt{Sarzi2018}) and provides the GC velocities in the inner regions of 32 Fornax cluster galaxies, many of them located in the cluster outskirts, and thus not shown in figure \ref{fig:los_map}, which displays the radial velocity of GCs from our sample within 1.5 square degrees.

The combined sample of GCs provides a representative probe of the whole GC system in the core of Fornax. The GC distribution in the innermost one square degree around NGC\,1399 is very uniform and geometrically complete. It amounts to more than 50\% of our total GC sample. 
To better visualize and identify patterns in the velocity distribution, we smooth the radial velocity with the locally weighted regression method LOESS \citep{Cleveland1988}. We implemented it with the python version developed by \cite{Cappellari2013}. LOESS tries to estimate the mean pattern by averaging the data into smaller bins. Normally, a linear or quadratic order polynomial is used in the LOESS technique. In our sample, some of the GCs in the phase space distribution were utterly isolated. Using a lower order polynomial could cause over-smoothing of distinct kinematic features. To prevent this, we used a 3$^{rd}$ order polynomial and a low value of smoothing factor of 0.1 \citep{Cleveland1979, Cleveland1988}. The LOESS smooth radial velocity map is shown in the right panel of figure \ref{fig:los_map}. 

\begin{figure*}
\centering
\includegraphics[width=1.0\linewidth]{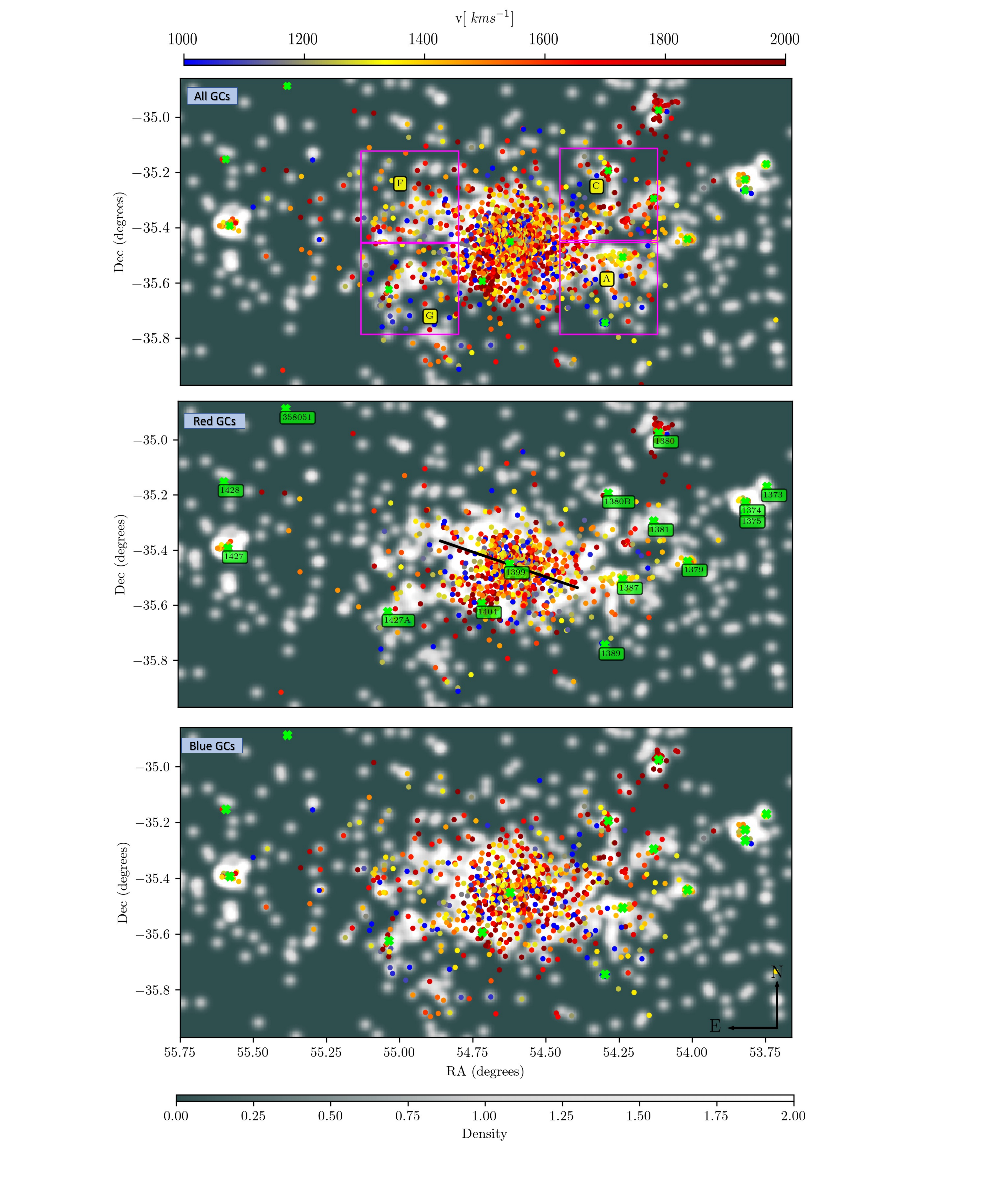}
\caption{GCs with confirmed radial velocities (coloured dots) are plotted over the surface density map of photometric GC candidates from the FDS \citep{Cantiello2020}. Top panel: Shown is the full GC radial velocity sample. Middle panel: Radial velocity distribution of red GCs. The black line shows the rotation axis of the red GCs at PA$=$70$^\circ$, measured north to east. Major galaxies are labelled in green. Bottom panel: The same for blue GCs. The density scale plotted on the bottom represents the number of GCs from the photometric sample per square arcminute.}
\label{fig:den_maps}
\end{figure*}

To check for any rotational signature in the GC system, we model the GC kinematics with a simple model that describes the rotational amplitude and velocity dispersion, similar to the work of \cite{Katja2020} (see their section 4.2.1). To have a homogeneous phase-space distribution of GCs, we consider GCs within 30 arcminutes from the central galaxy NGC 1399. In figure \ref{fig:vrot} we show the rotational amplitude and measured velocity dispersion for the full sample as well as red and blue GCs. We find a small rotational velocity of less than 30 \kms\ for the full sample and the blue GCs. Red GCs show a significant rotational velocity of 60 \kms\ with a rotation axis of PA$=$70$^\circ$ (measured North over East). The rotation axis for the entire sample is close to the one of red GCs and is certainly dominated by them. In figure \ref{fig:den_maps}, we show the rotation axis (black line) of the red GCs. For the red GCs, the ratio of $V_{rot}/\sigma$ is 0.22, meaning a low but significant rotational signature, similar to that of other massive galaxies. 

Only a few studies have examined the kinematics of the stellar body of NGC\,1399; for example, \cite{Saglia2000} used longslit observations to obtain the stellar kinematics out to 1.6' of NGC\,1399, giving a value of stellar $V_{rot}/\sigma$ $\sim$ 0.11. This low rotation measure is consistent with NGC\,1399's nearly round shape ($\varepsilon=0.1$), characterising this galaxy as a slow rotator. The central stellar kinematics of NGC\,1399 cannot be straightforward compared to the outer GC kinematics and rotation. They most probably reflect the formation of the central stellar spheroid via violent relaxation of early mergers. The photometric PA of NGC\,1399's major axis is $\sim$ 112$^\circ$ degrees (within 1') and varies between 90-110$^\circ$ at outer radii (up to 20') \citep{Iodice2016}, which is more or less consistent with the East-West elongation of the extended GC system. More measurements of the outer stellar kinematics around NGC\,1399 are needed to understand if the GCs and stellar halo components are kinematically coupled or decoupled.

Previously, \cite{Schuberth2010} studied the rotation of GCs around NGC 1399 and found a rotation amplitude of 61$\pm$35 km/s for the red GCs and 110-126 km/s for the blue GCs. In contrast to \cite{Schuberth2010}, we did not find a strong rotational signature for the blue GCs. This might be due to our large and uniform sample of GCs, whereas the \cite{Schuberth2010} sample was limited to 10 arcminutes and geometrically not complete.

We also notice different patches of low- ($<$1000\kms) and high-velocity regions ($>$1700\kms), elongated in east to west and north-east to south-west structures. We discuss the correlations between the photometric and kinematical properties of the GCs in the subsequent discussion section.

%--------------------------------------------------------------------

\begin{figure*}[h]
\centering
\includegraphics[width=0.85\linewidth]{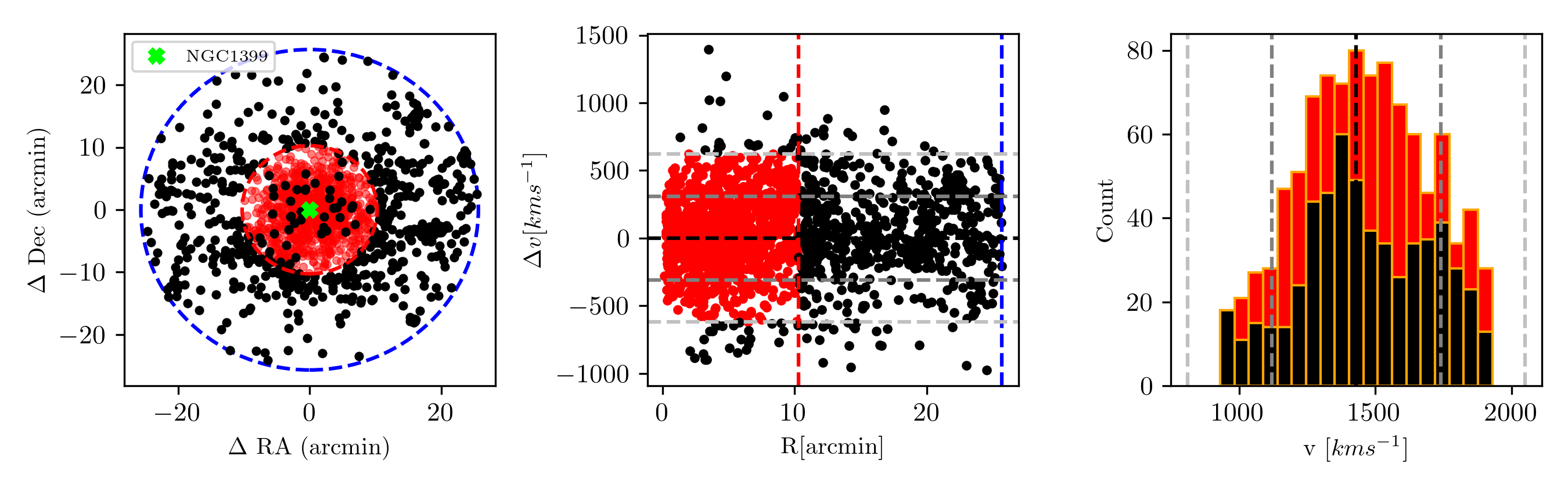}
\includegraphics[width=0.85\linewidth]{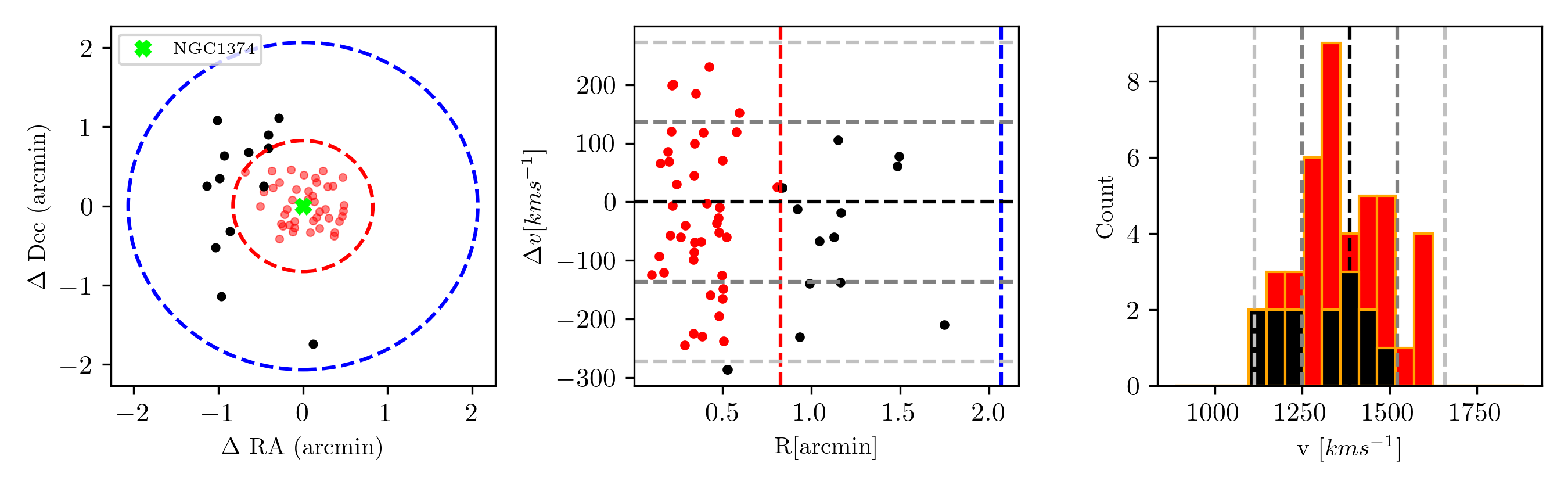}
\caption{Selection of potential ICGCs around NGC\,1399 (first row) and NGC\,1374 (second row). Left panel: the GC distribution within 5$r_{\rm eff}$ (blue dashed circle). The radius of 2$r_{\rm eff}$ is indicated as red dashed circle. Middle panel: the distribution of GCs in projected phase space. Red dots show the galactic GCs within 2$r_{\rm eff}$ and black dots the defined ICGCs. Vertical red and blue dashed lines indicate 2$r_{\rm eff}$ and 5$r_{\rm eff}$, and the grey horizontal lines give the 1- and 2-sigma scatter of GC velocities within 2$r_{\rm eff}$ around the systemic velocity of NGC\,1374. Right panel: Velocity histograms of of ICGCs (black) and galactic GCs (red) are shown. The black dashed line marks the LOS velocity of NGC\,1374 and the dashed grey lines the 1- and 2-$\sigma$ scatter of GCs radial velocities as also shown in the middle panel.}
\label{fig:ICGC_1374}
\end{figure*}

\begin{figure}[h]
\centering
\includegraphics[width=0.90\linewidth]{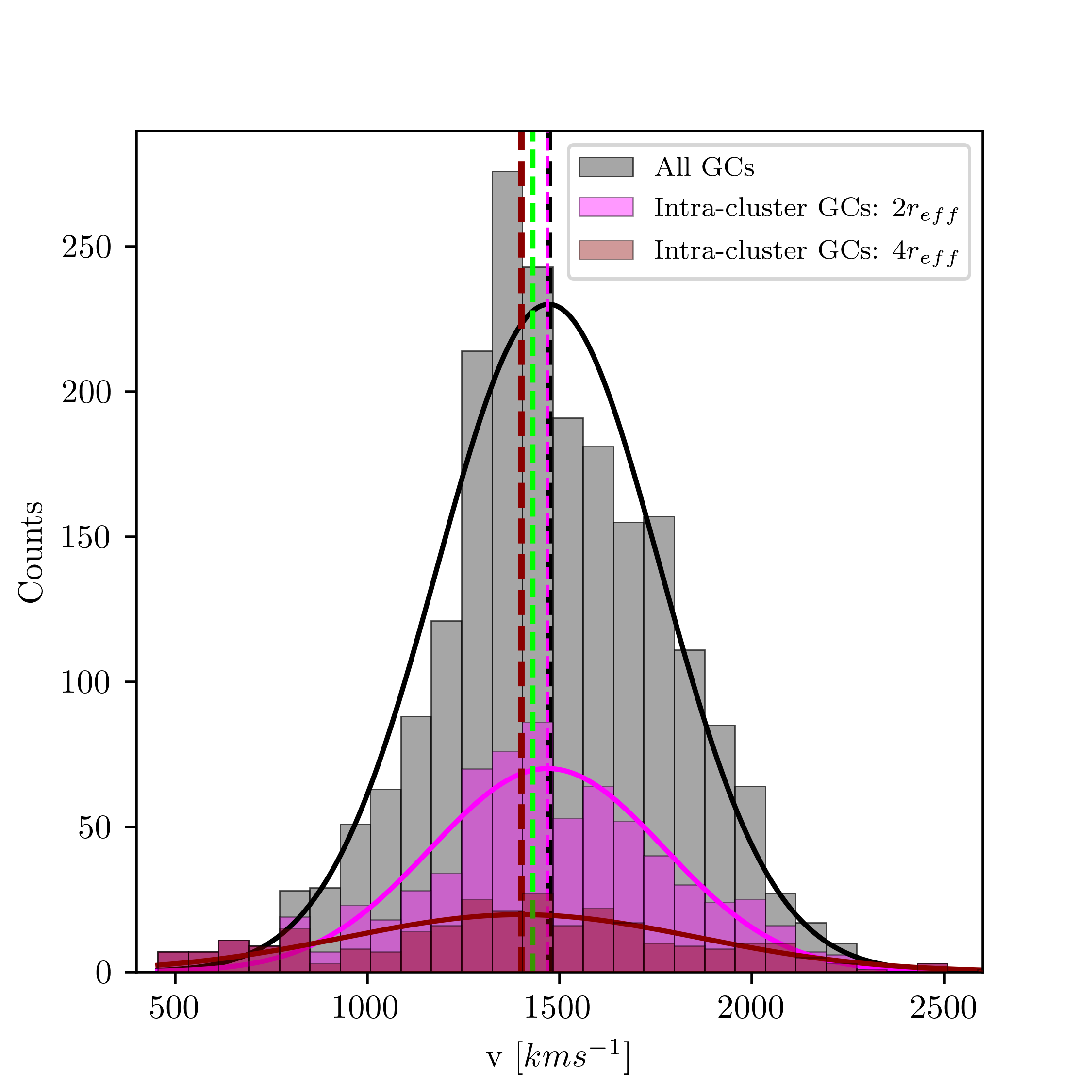}
\caption{Radial velocity histograms of the full sample and the intra-cluster GCs. The vertical green dashed line indicates the radial velocity of NGC\,1399. The vertical black, magenta, and dark red dashed lines indicate the mean velocities of fitted Gaussian to the three sets of GCs.}
\label{fig:all_vel_dis}
\end{figure}

\section{Discussion}\label{sec5}

In this section, we connect the photometrically discovered intra-cluster GCs with the full sample of 2341 confirmed GCs and study their phase-space distribution and radial velocity dispersion profile.

\subsection{Colours, phase space and spatial distribution}

We study the properties of red and blue GCs separately.
To divide the entire sample of 2341 GCs into red and blue sub-populations, we use the $g-i$ colour distribution since the shallower $u$-band photometry does not exist for all GCs. We adopt a value of $g-i=0.978$, obtained from the GMM (fig. \ref{fig:red_blue_hist}, left panel) to separate the two sub-populations. The brightest compact objects of our catalogue are a mix of genuine massive globular clusters and stripped nuclei, dubbed as ultra-compact dwarf galaxies (UCDs) in the literature \citep{Hilker1999a, Drinkwater2000, Drinkwater2003}. For selecting UCDs, we use a magnitude cut of $m_{i}<20$ mag \citep{Mieske2002} and found a total of 72 UCDs. In figure \ref{fig:cm_cmd}, we show the distribution of GCs and UCDs in the magnitude, color ($g-i$ and $g-r$) and velocity spaces. 

As can be seen from those plots, the UCDs are, on average redder than the GCs. This confirms the 'blue tilt' of bright GCs and UCDs that was already found in photometric samples of rich GC systems \citep[e.g.][]{Dirsch2003, Mieske2010, Fensch2014}. There exist some very blue and very red GCs, with $(g-i)<0.6$ and $(g-i)>1.6$, respectively. While the blue GCs might be explained by young to intermediate ages, the very red colours point to either very metal-rich populations, dust obscuration, or blends in the photometry. Future investigations are needed to clarify their nature.

In figure \ref{fig:rad_vel_dis}, we show the radial velocity of red (left panel) and blue (right panel) GCs as a function of the cluster-centric distance. Major galaxies around NGC\,1399 are shown as lime crosses. We observe that most of the red GCs are centrally concentrated on the systematic velocities of these galaxies \citep[taken from][]{Iodice2019}. Within 50\,kpc from NGC\,1399, red GCs homogenously span a range of relative velocities of $\pm$500\kms, and further outside follow a wedge shaped structure to smaller relative velocities till 150\,kpc. \cite{Schuberth2010} had observed the wedge shaped feature of the red GCs confined within 50\,kpc (see their Fig.\,9, right panel). However, with the current larger sample of GCs, we notice that it extends out to larger distances.

Most major galaxies at cluster-centric distances larger than 160\,kpc have similar systemic velocities as NGC\,1399. An interesting exception at $\sim$220\,kpc distance is NGC\,1380, which has a high systemic velocity of $\sim$1800\kms. Red GCs with similar high velocities are scattered out to $\pm50$ kpc galactocentric distances in filamentary structures around this galaxy, possibly suggesting a disturbance of its halo. Despite most red GCs being concentrated around major galaxies, there exists a noteworthy number of red GCs that seems not to be related to any particular galaxy. Those are candidates of intra-cluster GCs.

In contrast to red GCs, blue GCs show a more complex and irregular pattern in the phase space diagram. In particular, between 60-150\,kpc, they extend to larger relative velocities and fill the intra-cluster regions between the galaxies. Apart from that, blue GCs occupy the outer halos of the major galaxies.

Our UCD sample shows a radial velocity distribution in between 750-2500 \kms, with a mean velocity close to the radial velocity of NGC\,1399, with a velocity scatter of 312 \kms, consistent with the fainter GCs.
There also exists a dozen of red and blue GCs at low radial velocities of $\sim$500\kms\ at cluster-centric distances between 10-220\,kpc. This was already noticed for the blue GCs by \cite{Richtler2004}. Due to their high relative velocity with respect to the Fornax cluster, exceeding 800\kms, they might constitute unbound GCs from galaxy encounters with highly radial orbits in the line-of-sight, or a sheet of foreground 'intra-space' GCs. 

\subsection{Velocity dispersion profile}\label{sec5:vel_dis}

The large spatial coverage of our sample enables us to measure the velocity dispersion profile of the GCs out to 300\,kpc. For this, we define circular bins such that each bin has 100 GCs, and measure the velocity dispersion as the standard deviation of radial velocities in that bin. The uncertainty on the velocity dispersion is determined through a bootstrap technique. In each bin, we measure the velocity dispersion 1000 times and take its scatter as uncertainty. For the total sample, we obtain 23 bins, where the outermost bin has only 41 GCs. We followed the same procedure for red and blue GCs separately, resulting in 10 and 20 bins, respectively.

In figure \ref{fig:dis_pro} we show the velocity dispersion profile of our GC sample. The black line indicates the dispersion measurement for the full sample and the grey band denotes its 1$\sigma$ uncertainty. Blue and red dots indicate the values for the blue and red GCs. The vertical grey dashed lines show the projected cluster-centric distances of NGC\,1404 and other major galaxies. For reference and comparison, we have included the velocity dispersion measurements from previous studies as well, as indicated in the legend and caption.

We have also measured the velocity dispersion profile of potential intra-cluster GCs (ICGCs) within the Fornax core region, shown as a magenta line in the lower panel of figure \ref{fig:dis_pro}, with the light pink band indicating its 1\,$\sigma$ uncertainty.
The selection of the ICGCs was made by excluding GCs around major galaxies by performing cuts in the phase-space distribution. First, we calculate the scatter in the radial velocities of GCs within 2 effective radii ($r_{\rm eff}$) of each galaxy, \cite[taken from][]{Iodice2019}. We use $\pm2\sigma$ of this velocity scatter around the galaxy's LOS velocity as the lower and upper boundary to select the GCs belonging to each galaxy. The remaining GCs are classified as ICGC candidates, being aware of the fact  that this selection might include outer halo GCs that probably are still bound to their parent halos (see below).

Figure \ref{fig:ICGC_1374} shows examples of the ICGC candidate selection for the central galaxy NGC\,1399 and the galaxy NGC\,1374.  For the central galaxy, we clearly see a fraction of GCs with high relative velocities lying inside 2$r_{\rm eff}$, which are identified as the ICGCs. The true central concentration of ICGCs is difficult to access since they might overlap in radial velocity with GCs of the central galaxy. We note that our ICGCs selection criteria only provides a rough separation between GCs bound to individual galaxies and those belonging to an unbound, or at least disturbed intra-cluster population. Bound GCs might actually reach out to larger effective radii, but also true ICGCs might be projected at similar velocities in front or behind a galaxy and thus are `hidden' from detection. Only a detailed dynamical analysis of the mass profile around each galaxy can provide a cleaner sample of ICGCs. This is beyond the scope of this paper.

In any case, according to our selection criteria, 719 GCs, almost 31\% of the total sample, are classified as ICGCs. This number probably is an upper limit of true ICGSs due to the above-mentioned limitations of our selection criteria. Also, geometrical incompleteness of GC velocities within 2-4 $r_{\rm eff}$ around the major galaxies plays a role. Whereas the central regions are covered by MUSE observations, and thus GC counts are complete \citep{Katja2020} there, the outer halo regions are not fully covered by the VIMOS pointings, as can be seen in the uneven distribution of outer GCs around NGC\,1374 (fig. \ref{fig:ICGC_1374}, left panel).  

To produce a cleaner ICGC sample, we performed a similar selection as mentioned above but with a phase space cut at $>$4$r_{\rm eff}$. This left us with a sample of only 286 ICGCs (12\% of the total sample). The velocity dispersion profile of this set of GCs is shown as the dark-red band in the lower panel of figure \ref{fig:dis_pro}. 

Figure \ref{fig:all_vel_dis} shows the velocity distribution of the full and ICGCs samples. For all three samples, the mean velocity lies close to the radial velocity of NGC\,1399. The velocity scatter of the full sample and ICGCs, selected at 2$r_{\rm eff}$, is close to 300 \kms, whereas ICGCs selected outside 4$r_{\rm eff}$ show a larger velocity scatter of 455 \kms\ around a mean velocity of 1400 \kms. 
In the following, we describe the features and irregularities noticed in the velocity dispersion profiles, starting from the center outwards:

1) Between 2 and 5 arcminutes: The dispersion profile takes a steep rise from 220 to 350\kms\ within 1\,$r_{\rm eff}$ of NGC\,1399. Mostly blue GCs contribute to this rise. It is consistent with the rise previously reported by \cite{Schuberth2010}.  Red GCs show a constant velocity dispersion of $\sim270$\kms\ within 1\,$r_{\rm eff}$ of NGC\,1399, in agreement with \cite{Schuberth2010} and \cite{Pota2018}. 

2) Between 5 and 10 arcminutes: The total dispersion profile flattens around a value of $\sim$300\kms. While the dispersion profile of the blue GCs decreases, the one of the red GCs rises. This rise is caused by the superposition of the GCs of NGC\,1404, which has a high systemic velocity of 1944\kms. Within these radii limits, \cite{Spiniello2018} have also reported a similar increase in the PNe velocity dispersion profile. The lower velocity dispersion of red GCs from \cite{Schuberth2010} can be explained by the increase of sample size in our study, which added several GCs with more extreme velocities. For the ICGCs velocity dispersion profile, we observe a value of $\sim500$\kms at 8 arcmin. This is artificial and caused by the exclusion of GCs within 2\,$r_{\rm eff}$ of NGC\,1399. Thus, we are left with GCs with extreme radial velocities, resulting in the large dispersion value.

3) Beyond 10 arcminutes: The dispersion profile remains flat around $\sim$300\kms\ until 18 arcmin, consistent with \cite{Pota2018} results. Further out, the GCs belonging to individual galaxies dominate the velocity dispersion values of all GCs and cause large variations from $<$200 to $>$300\kms.
After a steep decrease from 500 to 300\kms, the velocity dispersion profile of ICGCs behaves smoother with nearly constant values around 290\kms\ out to 40 arcmin. This value is relatively consistent with the velocity dispersion of cluster galaxies \cite{Drinkwater2000}. A similar trend with PNe kinematics has been observed by \cite{Spiniello2018} out to 30 arcmin.

\cite{Iodice2016}, using $g$-band light distribution around NGC\,1399, identified a physical break radius at 10 arcmin, separating the total light profile into a central spheroid light of NGC\,1399 and an outer exponential halo. The constant value and flattening of the ICGC dispersion profile beyond 12 arcmin gives the kinematical confirmation to this physical break radius. 

\begin{figure*}
\centering
\includegraphics[width=0.95\linewidth]{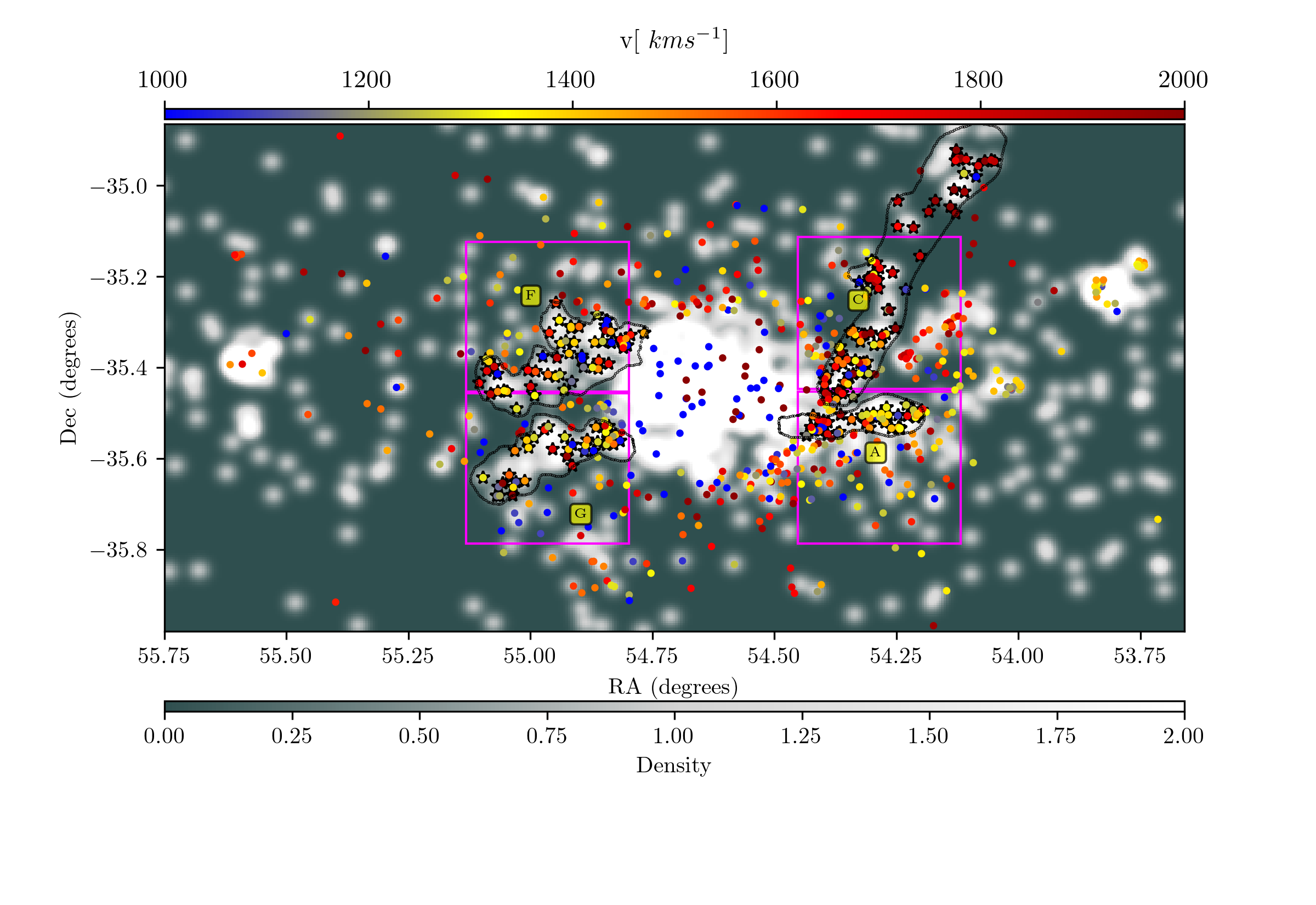}
\caption{Distribution of ICGCs selected with a phase space cut at $>$2$r_{\rm eff}$. GC overdensity regions are named as in \cite{Abrusco2016}. The magenta boxes show the regions where we perform 2D KS tests. Black contours show the visibly selected regions to study the stream properties.}
\label{fig:icgcs_map}
\end{figure*}

\begin{figure}[h]
\centering
\includegraphics[width=1.0\linewidth]{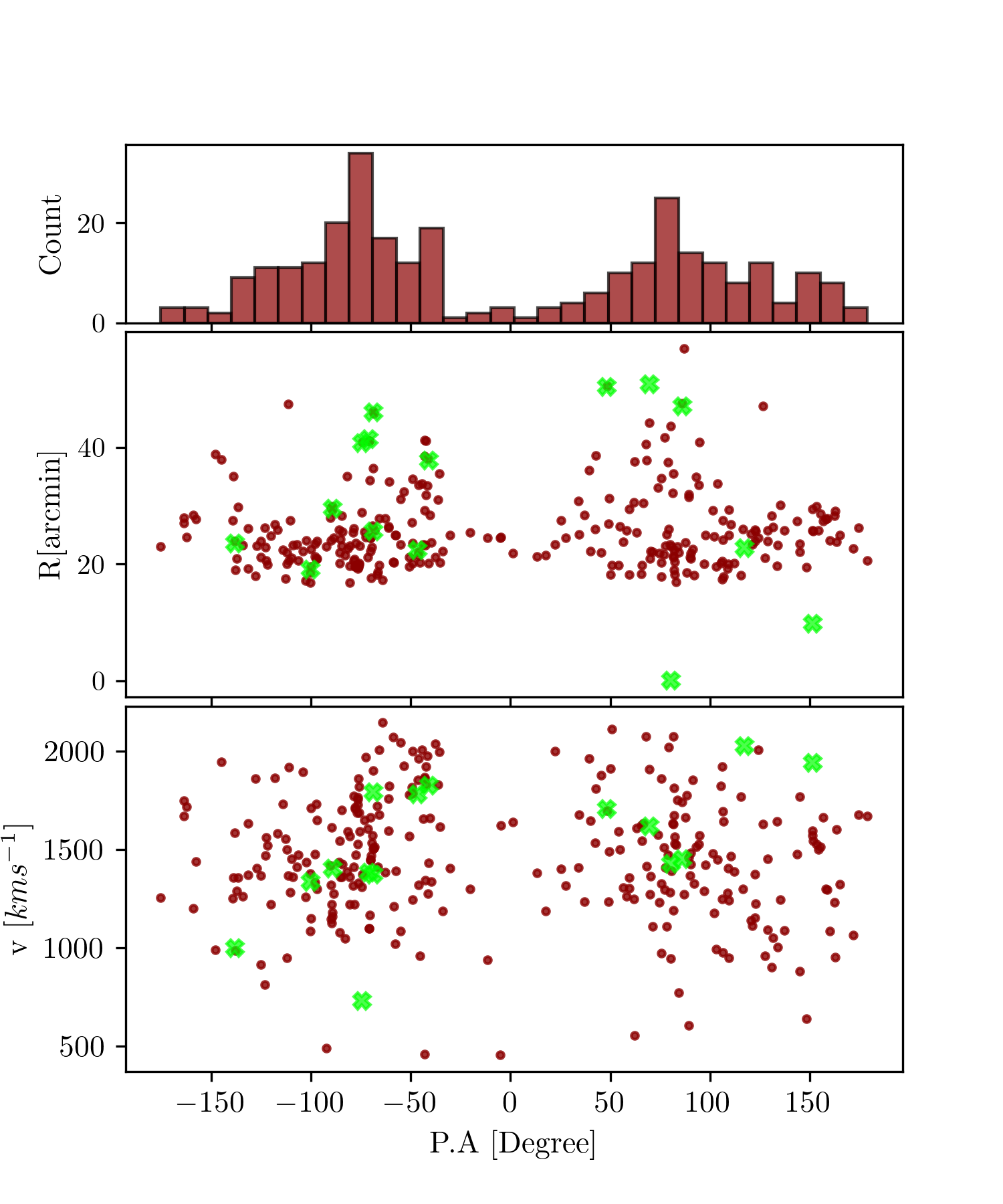}
\caption{Azimuthal distribution of ICGCs selected outside 4$r_{\rm eff}$ radii around major galaxies (see text for details). Top panel: Histogram of the ICGC position angle (PA), with bin size of 30$\circ$. Middle panel: Cluster-centric distance vs PA. Bottom panel: Radial velocity vs PA. Lime crosses indicate major galaxies.}
\label{fig:ICGC_PA}
\end{figure}

\subsection{Globular clusters and planetary nebulae}
\cite{Spiniello2018} presented the kinematics of 1452 PNe out to 200 kpc in the Fornax cluster core, spatially extending the results presented in \cite{McNeil2010}. Although the velocity dispersion profile of PNe overall follows the kinematics behaviour of the red GCs, slight differences in the profiles can be found. In figure \ref{fig:dis_pro}, the green dashed line shows the velocity dispersion profile of PNe taken from \cite{Spiniello2018}. Within 5 arcminutes, the velocity measurements we obtained for the red GCs show a slightly higher value than the PNe (this was true also in \cite{Spiniello2018}, but the difference between the velocity dispersion values was smaller). In between 5-10 arcminutes, the PNe show a high dispersion peak value at $\sim$380\kms, unlike our red GCs, and better match the value we measured for blue GCs. Between 10-20 arcmin, the velocity dispersion for both PNe and red GCs, decreases, with the red GCs showing a very low value at the projected distance of NGC\,1387. Beyond 20 arcmin, the PNe velocity dispersion shows a flat behaviour at $\sim$300\kms, slightly above that of blue and red GCs.
In general, the PNe velocity dispersion profile follows closely that of all GCs beyond 10 arcmin, rather than that of red or blue GCs individually. This might suggest that PNe trace the behaviour of both stellar populations, the one of galaxies as well as of intra-cluster light.

\subsection{Intracluster GC kinematics}

\begin{figure*}
\centering
\includegraphics[width=0.80\linewidth]{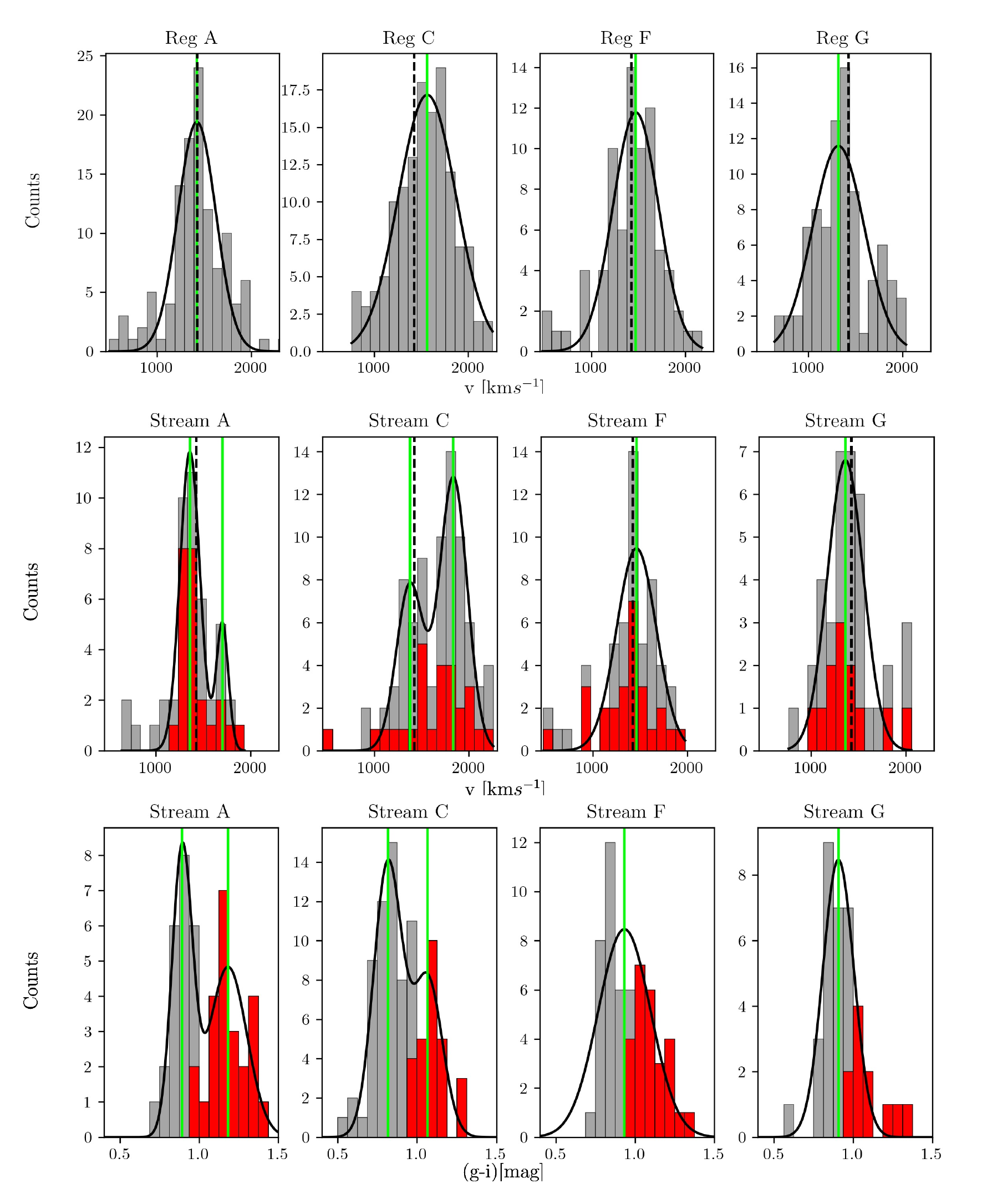}
\caption{Radial velocity and $g-i$ colour histograms of the intra-cluster regions. Top panel: radial velocity of GCs lying within the rectangular boxes A, C, F and G (left to right). Middle panel: radial velocity of GCs lying within the streams A, C, F, and G (left to right). The radial velocities of the red GCs are marked with red colour. Bottom panel: $g-i$ colour histogram of GCs lying within the streams A, C, F, and G (left to right). The black vertical dashed line marks the radial velocity of NGC\,1399, the vertical green lines show the peak positions of the fitted Gaussians. For streams A and C, double Gaussians were fitted.}
\label{fig:Intra_vel_hist}
\end{figure*}

\begin{table*}
\caption{2D KS test and properties of the ICGCs.}
\label{Tab3:2dks_test}
\centering
\begin{tabular}{l c c c c} 
\hline
Intra-cluster region & p-value & Median velocity [\kms] & Velocity scatter [\kms] & \ Blue to red GCs ratio  \\  
(1) & (2) & (3)& (4) & (5) \\  
        \hline
        Reg A     & 0.37 & 1419 & 200  & 1.17\\
        Reg C     & 0.72 & 1559 & 303  & 1.92\\
        Reg F     & 0.84 & 1477 & 312  & 1.56\\
        Reg G     & 0.66 & 1352 & 291  & 1.51\\
        Stream A     & 0.48 & 1375 & 127 & 0.87 \\
        Stream A $2^{nd}$ peak & -- & 1723 & 56 & -- \\
        Stream C     & 0.20 & 1384 & 117 & 2.03  \\
        Stream C $2^{nd}$ peak & -- & 1824 & 128 & -- \\
        Stream F     & 0.98 & 1452 & 228 & 1.15  \\
        Stream G     & 0.54 & 1362 & 166 & 2.08  \\
        \hline
  \end{tabular}
    \begin{tablenotes}
      \small
        \item \textbf{Notes:} Column 3 and 4 show the mean velocity and velocity scatter of GCs within respective intra-cluster region.
    \end{tablenotes}
\end{table*}

The first photometric wide-field search for GCs in the Fornax cluster by \cite{Bassino2006} reported the existence of an ICGC populations based on GC overdensities in regions between the central galaxy NGC\,1399 and neighbouring galaxies. Later, \cite{Bergond2007} and \cite{Schuberth2008} kinematically identified and quantified the properties of some ICGCs. Through the FDS survey, \cite{Abrusco2016} reported the discovery of an extended GC density distribution in the Fornax core region with several well defined overdense regions, and \cite{Iodice2016} discovered a faint stellar bridge coinciding with the GC over density between NGC\,1399 and NGC\,1387, confirming the interaction between these two galaxies. 

Our extended and spatially homogeneous GC catalogue allows us to study the kinematical properties of the enhanced density regions of GCs in Fornax. 
In fig.\,\ref{fig:den_maps}, we plot our full GC radial velocity sample on top of a smoothed density distribution of photometric GC candidates by \cite{Cantiello2020}. 
To create the smoothed density map, we use the non-parametric kernel density estimates based on python-scikit-learn kernel density routine by \cite{scikit-kde}. From the FDS catalogue, a density of ~0.75 GCs/arcmin$^2$ is expected within the central region of the Fornax cluster. To include at least a couple of GCs in the density maps and to create the visual impression of the GCs streams, we adopted a Gaussian kernel bandwidth of 0.015 degrees, which is $\sim$1 arcmin. The top, middle, and lower panels show the distribution of all, red and blue GC candidates, respectively.

Confirming the FDS survey findings of \cite{Abrusco2016} and \cite{Cantiello2020}, we also observe an elongated distribution of confirmed GCs in the east-west direction, centred on NGC\,1399. Looking at the radial velocity patterns in the smoothed velocity map in fig. \ref{fig:los_map}, we find that, on the west side of NGC\,1399, GCs have a relatively higher radial velocity than on the east side. Further on the east side of NGC\,1399, in the over-dense G and F features, GCs show an extended, filamentary spatial distribution. 

The azimuthal distribution of ICGCs selected outside 4$r_{\rm eff}$ of major galaxies is shown in figure \ref{fig:ICGC_PA}. It highlights the east-west elongation of ICGCs around NGC\,1399. In the two lower panels of that figure, where the azimuthal distribution is shown as a function of cluster-centric distance and radial velocity, respectively, phase-space features in between galaxies become apparent. A detailed dynamical analysis of those features is beyond the scope of this paper.

We investigated the spatial correlation between the photometric and our radial velocity catalogue. For this, we perform a 2-dimensional Kolmogorov-Smirnov (2D-KS) test \citep{Peacock1983} in the four GC overdensity regions named Reg. A, C, F and G following the same naming convention as in \cite{Abrusco2016}. Fig. \ref{fig:icgcs_map} shows the phase-space distribution of the ICGCs selected with a cut of $>$2$r_{\rm eff}$ and indicates the rectangular regions around the GC overdensities. For all the regions, we get a p-value higher than 0.20, which means that the spatial distributions of photometrically selected and confirmed GCs are correlated by more than $3\sigma$ significance. Figure \ref{fig:Intra_vel_hist} shows the radial velocity histograms of ICGCs falling within these four overdense regions. The mean velocities are $\sim$1450 \kms, close to the radial velocity of NGC\,1399 and the Fornax cluster itself. This suggests that the GCs in these regions are ICGCs, kinematically influenced by the Fornax cluster potential. In table \ref{Tab3:2dks_test}, we list the p-values obtained from the 2D-KS tests for all regions and the fitted Gaussian mean and velocity scatter. These selected rectangular regions inhabit both red and blue GC populations, but on average, the blue GCs dominate in numbers. In contrast to this, GCs within the 2$r_{\rm eff}$ radii of galaxies show, on average, a higher fraction of red GCs. We list the number ratio of blue to red GCs in table \ref{Tab3:2dks_test}.

With the small sample size of ICGCs, and given the caveats of the rough ICGC selection criteria mentioned in \ref{sec5:vel_dis}, it is hard to speculate about the nature of the GCs in the over-dense regions. Looking at figure \ref{fig:icgcs_map} it is quite clear that our spectroscopic sample provides kinematical information on the visible photometric streams in the overdense regions. Our 2D KS test performed for the selected rectangular regions, demonstrates that our spectroscopic sample is statistically coherent with the photometric sample. To get a hint on the possible progenitor galaxies of ICGCs and their physical properties, specifically, the ones which are tracing the visible streams (marked in figure \ref{fig:icgcs_map} with the black contours), we plot the radial velocity and $g-i$ colour histograms in figure \ref{fig:Intra_vel_hist}. We name these streams A, C, F and G. In the following we explain the features noticed within these streams:

Stream A is a feature related to the faint stellar bridge reported by \cite{Iodice2016}, connecting NGC\,1387 and NGC\,1399 (region A).
In this stream, the GCs radial velocity distribution shows two peaks, one close to the radial velocity of the central galaxy NGC\,1399 (at 1374 \kms) and the other at 1723 \kms. Both peaks show a low velocity scatter of values 127 \kms\ and 56 \kms, respectively. The second peak is possibly arising from GCs on the east side of NGC\,1387, which is interacting with NGC\,1399. These GCs might be tidally stripped off the halo of NGC\,1387. In the radial velocity histograms, we mark the contribution of red GCs in red colour. We notice that red GCs mostly contribute to the first radial velocity peak, whereas the blue GCs dominate the second peak. This suggests that tidally-stipped GCs, those that are outside the systemic Fornax cluster velocity, are mostly blue. In the $g-i$ colour histogram of stream A (bottom panel of fig. \ref{fig:Intra_vel_hist}), we also observe two peaks, suggesting the presence of red as well as blue GCs. Studying the GC colour distributions of early-type galaxies in the Virgo cluster \cite{Peng2006} have shown that luminous galaxies ($M_{B}$ $\sim$ -21 mag) have mostly bimodal GC colour distributions, whereas low luminosity galaxies ($M_{B}\sim-16$) have dominant fractions of blue GCs. In stream A, we see a bimodality in the $g-i$ colour histogram with a larger fraction of red GCs. Comparing this kind of bimodality with results of \cite{Peng2006} suggests that GCs in streams A are mostly generated by the interaction of the luminous galaxy NGC\,1387 and NGC\,1399.

Stream C in the overdense region C, consists of a chain of GCs in the vicinity of NGC\,1380 and NGC\,1380B. \cite{Cantiello2020} pointed out that this GC overdensity could result from the LOS projection of adjacent GC systems. Indeed, the GCs in the vicinity of NGC\,1380 and NGC\,1380B (but beyond 2$r_{\rm eff}$ radii), have radial velocities larger than 1700 \kms, consistent with the systemic velocities of both galaxies. In the GC radial velocity histogram of stream C we also observe two peaks, one close to the radial velocity of NGC\,1380 and the other close to the NGC\,1399 radial velocity. Similar to stream A, stream C GCs also show a bimodal $g-i$ colour histogram, with a higher fraction of the blue GC population, although the bimodality is not as clear due to the small sample size. In stream C, both radial velocity peaks are dominated by blue GCs with radial velocity scatters of $\sim$110 \kms. In stream C, both radial velocity peaks are dominated by the blue GCs with radial velocity scatters of around ~110 km/s. In stream C, blue GCs are almost twice as abundant than red GCs. As shown in the studies of \cite{Peng2006}, an asymmetrical distribution of GCs, with an inclination towards blue GCs, suggests that the GCs in stream C are generated by galaxies in the magnitude range $-20<M_{B}<-19$ mag.

In stream F, GCs show a radial velocity distribution in between 700-1800 \kms with a mean velocity and scatter of 1452 \kms\ and 228 \kms, respectively. We find an equal fraction of blue to red GCs in stream F.

Stream G harbours GCs in the velocity range 800-2000 \kms, with a peak velocity close to the systemic velocity of Fornax and is mostly dominated by the blue GCs. Those GCs comprise a kinematically coherent group and in the $g-i$ colour histogram, we observe a peak towards blue GCs, suggesting that the progenitors of GCs in stream G are low luminosity galaxies \citep{Peng2006}.

Wee also notice that in the south-east of NGC\,1404, next to the G feature, some GCs show radial velocities higher than 1700\kms, similar to the systemic velocity of NGC\,1404. As previously shown by \cite{Bekki2003}, and more recently through X-ray studies by \cite{Su2017}, NGC\,1404 suffered from tidal interaction with NGC\,1399 in the past few gigayears. Thus, tidally released GCs are expected around NGC\,1404, and we now might see for the first time, the kinematical signature of those. Detailed dynamical models will be necessary to assess which GCs in the overall phase space distribution might have belonged to NGC\,1404 in the past.

Furthermore, we notice that GCs around NGC\,1427A, on the south-west side of NGC\,1399, show a stream-like distribution with a gradual decrease in radial velocity from 1500\kms\ north of \,1427A to 1100\kms\ south of the galaxy. This kinematic feature might suggest that NGC\,1427A is moving in south-north direction, loosing its GCs during its cruise through the core of the Fornax cluster \citep[e.g.][]{Lee2018}. 

Finally, we looked for the spatial distribution and numbers of red and blue GC sub-populations in the over-dense regions. As can be seen in figure \ref{fig:den_maps}, blue GCs dominate the intra-cluster over-dense regions between the Fornax cluster galaxies, whereas red GCs are more concentrated on the galaxies. The dominance of the blue (and thus mostly metal-poor) GCs in the Fornax IC regions and in the visually identified streams suggests that the Fornax IC component results from the accretion of tidally stripped low mass galaxies. Our results are in accordance with the dominantly blue ICGCs population observed for the Virgo cluster \citep[see][]{Ko2017, Longobardiicgs2018}. We list the number ratio of blue to red GCs in each region and the fours streams in table 3.

%--------------------------------------------------------------------
\section{Conclusions}\label{sec6}

We have re-analyzed VLT/VIMOS data of the central one square degree of the Fornax cluster, leading to produce radial velocity measurements of 777 GCs, which we present in a catalogue. Adding literature data, this provided us the largest and spatially most extended compilation of GC radial velocities in the Fornax cluster. This sample was used to kinematically characterize GCs in the core of the cluster. In the following, we highlight the main results of our work:

1) With the improved VIMOS ESO reflex pipeline 3.3.0  and careful analysis of radial velocity measurements with pPXF over the full spectral range, we have doubled the number of GC radial velocity measurements on the same dataset that was previously analyzed by \cite{Pota2018}. Combined with previously measured values from the literature, we gathered a sample of 2341 GC radial velocities in Fornax. 

2) We used the Gaussian mixture modelling technique to divide the full sample of 2341 GCs into a blue (56\%) and a red (44\%) GCs sub-population. The phase space distribution of red GCs shows that most of them are bound to the major cluster galaxies, in particular the central galaxy NGC\,1399. In contrast, blue GCs are spatially extended and show more irregular kinematics patterns. They occupy the outer haloes of galaxies and the intra-cluster space.

3) Using the radial velocities of GCs, we measured the dispersion profile out to a radius of 300\,kpc, covering almost half of the virial radius of the Fornax cluster. Beyond 10 arcmin ($\sim$58\,kpc), the dispersion profile of all GCs flattens. This radius is therefore considered as the break radius separating the potential of NGC\,1399 from that of the cluster. This result is strongly confirmed by the dispersion profile of potential ICGCs, which shows a flat behaviour beyond 10 arcmin at a value of 300$\pm$50\kms.

4) The radial velocity map of the full GCs sample kinematically characterizes the previously photometrically discovered ICGC population of the Fornax cluster. The different over-dense GC regions are marked by streams of higher relative velocity GCs, giving first kinematical evidence of interactions between the central galaxy NGC\,1399 and other major galaxies. 

5) Finally, we notice that mostly blue GCs dominate the intra-cluster regions and trace sub-structures that connect NGC\,1399 to its neighbouring galaxies.

With the future goal to study the Fornax cluster's mass distribution and assembly history, the presented GC radial velocity catalogue is of unprecedented value in exploring the dynamical structure and evolution of the Fornax cluster and its member galaxies.

\small
\textbf{Acknowledgements:} Our sincere thanks to the anonymous referee for helpful feedback and suggestions that improved the manuscript's scientific content. The bulk of the velocities were derived from the FVSS data taken under the ESO programme 094.B-0687 (PI: Capaccioli). Our special thanks goes to Massimo Cappacioli who dedicated INAF GTO time to spectroscopic Fornax cluster projects. Some velocities of our full catalogue (including literature data) are based on so far unpublished FORS2 data taken under ESO programmes 078.B-0632 and 080.B-0337 (PI: Hilker). A. Chaturvedi acknowledges the support from the IMPRS on Astrophysics at the ESO and LMU Munich. A. Chaturvedi also thanks Lodovico Coccato for helpful discussions. M. Cantiello acknowledges support from MIUR, PRIN 2017 (grant 20179ZF5KS). N.R. Napolitano acknowledges financial support from the “One hundred top talent program of Sun Yat-sen University” grant N. 71000-18841229, and from the European Union Horizon 2020 research and innovation programme under the Marie Skodowska-Curie grant agreement n. 721463 to the SUNDIAL ITN network. G.v.d. Ven acknowledges funding from the European Research Council (ERC) under the European Union's Horizon 2020 research and innovation programme under grant agreement No 724857 (Consolidator Grant ArcheoDyn). C. Spiniello is supported by a Hintze Fellowship at the Oxford Centre for Astrophysical Surveys, which is funded through generous support from the Hintze Family Charitable Foundation. This research made use of Astropy (\href{https://www.astropy.org/}{https://www.astropy.org})- a community-developed core Python package for Astronomy \citep{Astropy2013, Astropy2018}.

\bibliographystyle{aa} % style aa.bst

\bibliography{paper1_catalog}

\begin{appendix}
\onecolumn
\section{Test regarding pPXF parameter choices} \label{polychoise}

\textbf{Effect of initial velocity guesses on the resulting pPXF radial velocity: }
To check how much the pPXF initial radial velocity guess can affect the resulting radial velocity, we fitted a few spectra with varying initial velocity guesses ranging between 800 and 1800 \kms\ with a step of 200 \kms. The figure shows three example spectra at S/N$=$6, 12 and 17. 
The resulting radial velocities are pretty consistent, and the variations are less than 5\%, which is smaller than the individual velocity measurement errors.

\begin{figure*}[h]
\center
\includegraphics[width=0.35
\linewidth]{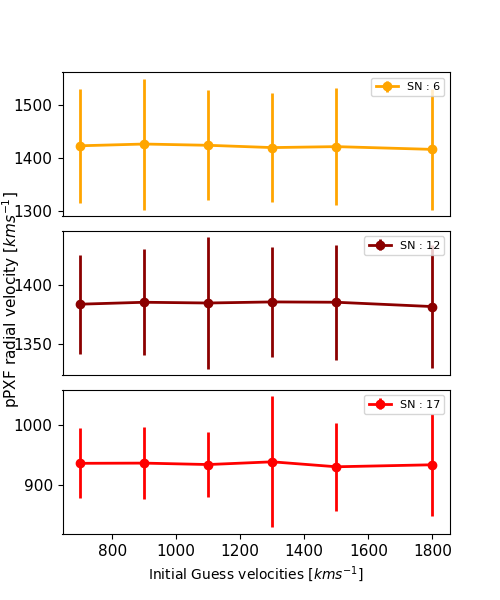}
\caption{\small{Effect of initial radial velocity guesses on the resulting radial velocity measurements.}}
\label{fig:vel_guess}
\end{figure*}

\begin{figure*}[h]
\center
\includegraphics[width=0.85\linewidth]{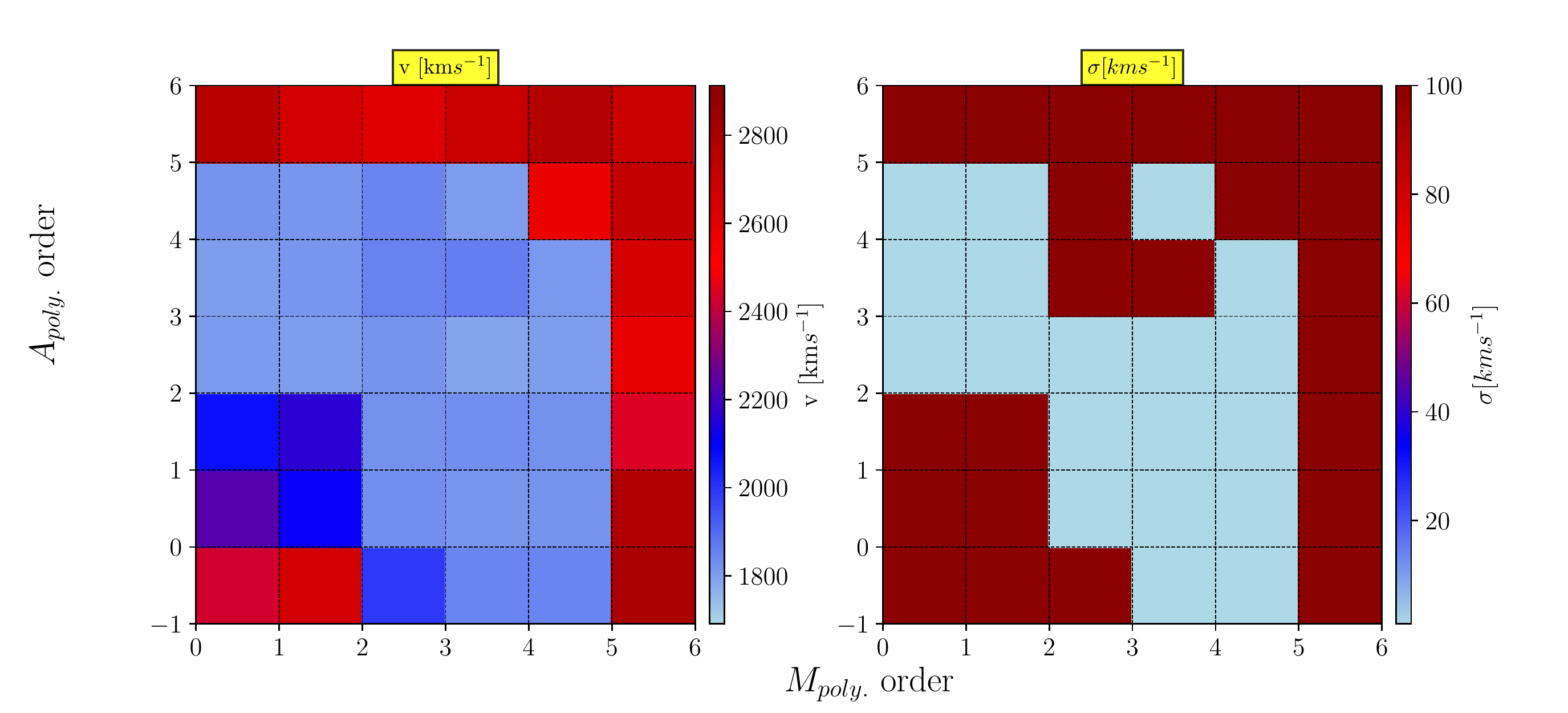}
\caption{\small{Choosing pPXF polynomials. Derived radial velocity (left panel) and the velocity dispersion (right panel) for a GC spectrum of S/N$\sim5$ as a function of additive and multiplicative polynomial degree. Assigning to the additive polynomial a grade equal to -1 means running pPXF without it, i.e. using only multiplicative polynomials.}}
\label{fig:poly_opt}
\end{figure*}

\textbf{Variation in the derived radial velocity and velocity dispersion as a function of the polynomial degree:} Here we present the results of tests mentioned in sect.\,\ref{ppxf_los}. For GCs where we obtain a radial velocity consistent with the Fornax cluster ($450<v<2500$\kms), but a velocity dispersion higher than 20 \kms, we varied the additive and multiplicative polynomials to quantify the effect on the derived radial velocity and velocity dispersion. Based on consistent values of radial velocity and dispersion we selected the additive and multiplicative polynomials. In figure \ref{fig:poly_opt} we show the radial velocity (left) and velocity dispersion (right) measured for one spectrum, on a grid of additive and multiplicative polynomials with orders in between 0 and 6. As can be noticed from these plots, the radial velocity remains quite consistent for additive polynomials with orders of 2 to 5 and multiplicative polynomial with orders in between 2 and 5. At the same time, the velocity dispersion remains lower than 20\kms\ within this polynomial range.

\begin{figure*}[h]
\center
\includegraphics[width=0.35
\linewidth]{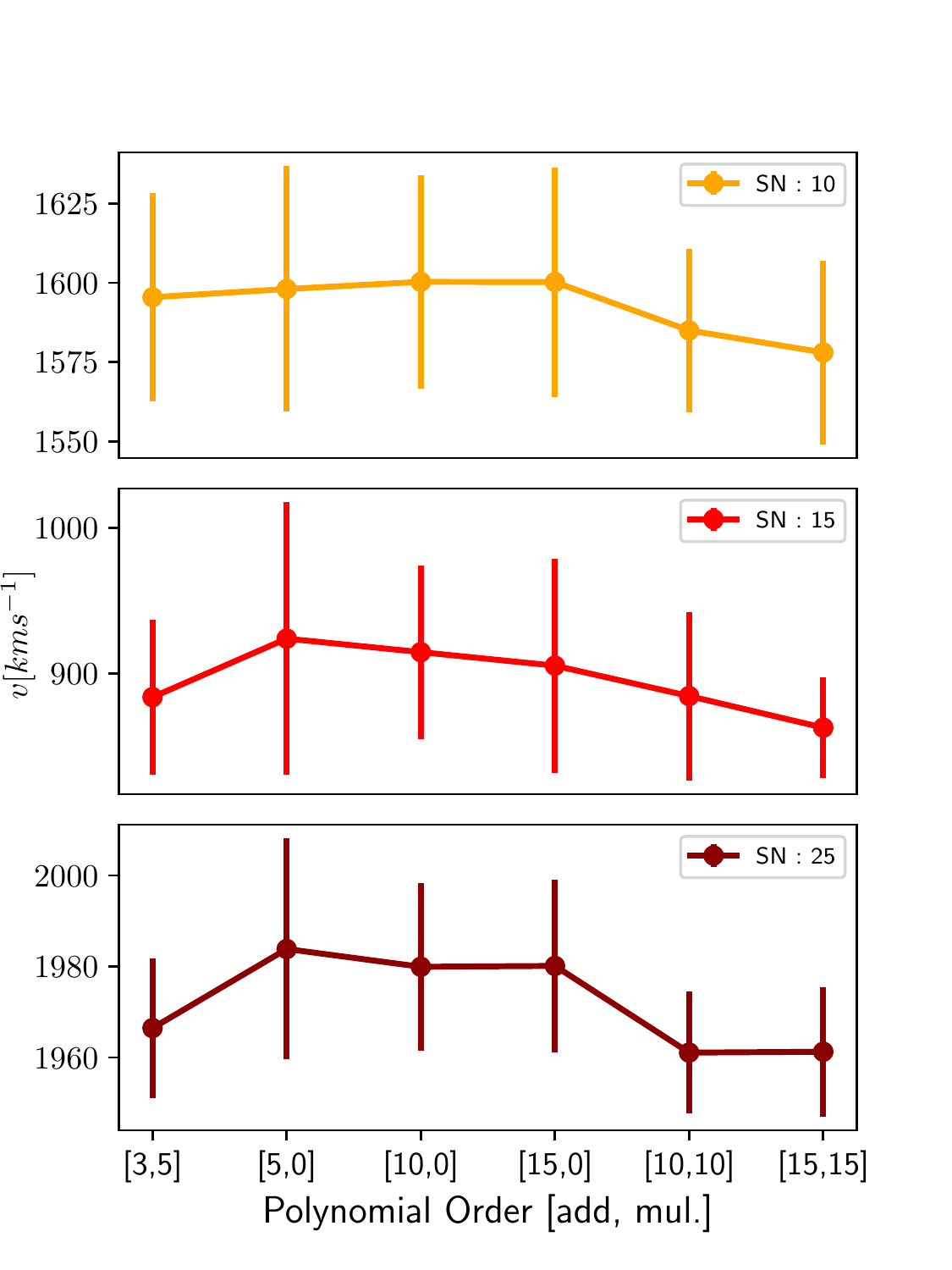}
\caption{Effect of higher order polynomials on radial velocity measurements. }
\label{fig:poly_high}
\end{figure*}

\textbf{Effect of higher order pPXF polynomials on radial velocity: }
To quantify the effect of higher order polynomials on the inferred radial velocities, as well as using only additive polynomials, as often done when fitting stellar kinematics, we used five sets of additive and multiplicative polynomials denoted as [additive, multiplicative] : [3,5] (our choice), [5,0], [10,0], [15,0], [10,10], [15,15]. The measured radial velocities from pPXF for these sets of polynomials for spectra of three different S/N are shown in figure \ref{fig:poly_high}. The derived radial velocities are always consistent and show only slight variations, which are within 5\%.

\newpage
\section{Example of a GC portfolio}\label{gc_portfolio}
\begin{figure*}[h]
\center
\includegraphics[width=0.85\linewidth]{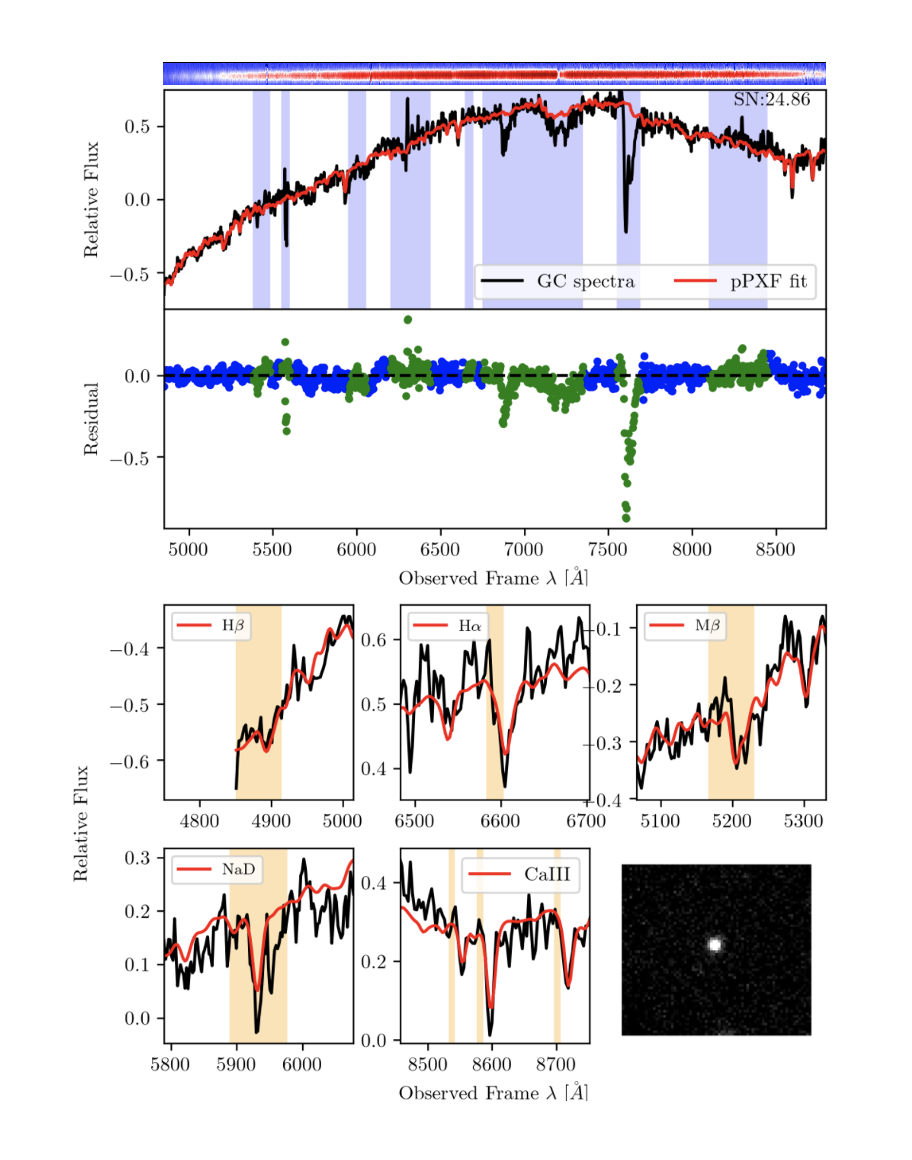}
\caption{Example of a GCs portfolio. The top panel shows the 1-dimensional extracted spectrum from the 2-dimensional VIMOS observation (on top). The middle panel shows the pPXF fit and its residuals in the subplot. The lower panel shows zoom-in wavelengths regions of absorption features used for visually selecting bonafide GCs. The lower right sub-panel shows the 2d image of the GC from pre-imaging. }
\label{fig:gcs_portfolio}
\end{figure*}

\newpage

\section{Catalog of VIMOS dataset
GCs}\label{gc_catalog}

Table \ref{gc_catalog} shows an overview of our VIMOS data GCs radial velocity catalog. Together with the A and B class objects, we have also included class C objects in the catalogue, totalling 851 GCs (see the sect. \ref{sec4} for details.) For each GC, we provide its measured radial velocity and its uncertainty, S/N, and object class (A, B and C). We also give magnitude information for each GC in $g$,$r$, $i$, and $u$ band obtained from the matched FDS photometric catalogue presented by \cite{Cantiello2020}.

\begin{table}[H]
    %\centering
    \caption{Excerpt of the present VIMOS GCs catalogue.}
    \begin{tabular}{%
    cccccccc}
    \toprule
    \toprule
    {Point Name} & {FVSS-GC-ID} & {R.A. (J2000)} & {Dec.(J2000)} & {v [\kms]} &  {$\Delta$ v[\kms]} & {S/N} & {Object Class} \\
    {(1)} & {(2)} & {(3)} &  {(4)} &  {(5)} & {(6)} & {(7)} & {(8)}\\
    \midrule
    \midrule
    fnx01\_q1\_ext1\_4\_slit\_73 & FVSS-GC:54.76610:-35.40796 & 54.766100 & -35.407960 &       1948.80 &     14.74 &  25.39 &          A \\
   fnx01\_q1\_ext1\_9\_slit\_214 & FVSS-GC:54.74755:-35.35828 & 54.747550 & -35.358280 &        799.97 &     88.59 &  15.03 &          A \\
   fnx01\_q1\_ext1\_10\_slit\_201 & FVSS-GC:54.74082:-35.35343 & 54.740820 & -35.353430 &       1675.47 &    152.62 &  13.57 &          A \\  
    fnx01\_q1\_ext1\_12\_slit\_59 & FVSS-GC:54.73409:-35.41361 & 54.734090 & -35.413610 &        681.64 &     59.39 &  15.45 &          A \\

    \bottomrule
    \end{tabular}
\end{table}

\addtocounter{table}{-1}
\begin{table}[H]
    \center
    \begin{tabular}{%
    ccccccc}
    \toprule
    \toprule
    {ID (FDS) } & {R.A. (J2000) (FDS)} & {Dec. (J2000) (FDS)} & {g[mag]} & {$\Delta$ g[mag]} &     {r[mag]} & {$\Delta$ r[mag]} \\
    {(9)} & {(10)} & {(11)} &  {(12)} &  {(13)} & {(14)} & {(15)} \\
    \midrule
    \midrule
    FDSJ033903.86-352428.64 &  54.766102 &   -35.407955 &              21.704 &  0.014 & 20.959 &  0.014   \\
    FDSJ033859.41-352129.71 &   54.747524 &  -35.358253 &  
    22.223 & 0.015  & 21.53 & 0.017   \\
    FDSJ033857.80-352112.32 &   54.740826 & -35.353424 &              22.244 &  0.017 & 21.636 & 0.016 \\ 
    FDSJ033856.18-352448.90 &  54.734097 &  -35.413582 &              20.651 &  0.016 & 19.958 &  0.013 \\ 

    \bottomrule
   \end{tabular}
\end{table}

\addtocounter{table}{-1}
\begin{table}[H]
    \center
    %\caption{Table 1}
    \begin{tabular}{%
    cccc}
    \toprule
    \toprule
    {i[mag]} & {$\Delta$ i[mag]} &  {u[mag]} & {$\Delta$ u[mag]} \\
    {(16)} & {(17)} & {(18)} &  {(19)}  \\
    \midrule
    \midrule
20.566 &  0.013 &  23.558 & 0.094 \\
21.530 &  0.017 &  21.287 & 0.087 \\
21.375 &  0.024 &  23.276 & 0.074 \\
19.707 &  0.012 &  21.975 & 0.031 \\

    \bottomrule
   \end{tabular}
\end{table}

\textbf{Notes.} Column list:(1)  GC named as VIMOS pointing id; (2) CGs named as FVSS ID (FVSS-GC:RA:DEC); (3)Right ascension;
(4) Declination; (5) GC Radial velocity; (6) Radial velocity uncertainty;
(7) Spectral S/N; (8) GCs object class; (9) FDS ID (10) Right ascension (FDS) ;
(11) Declination (FDS); (12-13) g band magnitude with error;
(14-15) r band magnitude and its error;
(16-17) i band magnitude and its error;
(18-19) u band magnitude and its error

\end{appendix}
\end{document}